\documentclass[10pt,a4paper]{article}  
\usepackage{multicol}
\usepackage{graphicx}
\usepackage{color}
\usepackage[colorlinks,citecolor=blue,urlcolor=blue,linkcolor=blue]{hyperref}
\usepackage{amssymb}
\usepackage{amsmath}
\usepackage{cite}
\newcommand{\Ppa}{P_\parallel}

\newcommand{\Qpa}{Q_\parallel}
\newcommand{\Pupa}{P_{1\parallel}}
\newcommand{\Qupa}{Q_{1\parallel}}
\newcommand{\Ppe}{P_\bot}

\newcommand{\Qpe}{Q_\bot}
\newcommand{\Rpapa}{R_{\parallel\parallel}}
\newcommand{\Rpape}{R_{\parallel\bot}}
\newcommand{\lb}{\frac{\omega}{k}-v}
\newcommand{\tgh}{\mathrm{th}}
\newcommand{\gm}{}
\newcommand{\poto}{\rule[-0.7em]{0pt}{2em}}
\newcommand{\buildlow}[2]{\mathrel{\mathop{\kern 0pt#1}\limits_{#2}}}
\begin{document}

{\Large\textbf{An extended hydrodynamics model for inertial confinement fusion hohlraums}}
\vskip 1em
\textbf{O. Larroche}\footnote{e-mail: olivier.larroche@cea.fr}$^{,1,2}$
\vskip 1em
$^1$CEA DAM DIF, 91297 Arpajon Cedex, France

$^2$Universit\'e Paris-Saclay, CEA, LMCE, 91680 Bruy\`eres-le-Ch\^atel, France
\vskip 2em
This preprint has not undergone peer review or any post-submission improvements or corrections. The Version of Record of this article is published in the European Physical Journal D, and is available online at https://doi.org/10.1140/epjd/s10053-021-00305-2 .
\vskip 2em
{\Large\textbf{Abstract}}
\vskip 1em
In some inertial confinement fusion hohlraum designs, the inside plasma is not sufficiently collisional to be satisfactorily described by the Euler equations implemented in hydrodynamic simulation codes, particularly in converging regions of the expanding plasma flow.
To better treat that situation, this paper presents an extended hydrodynamics model including higher moments of the particle velocity distribution function, together with physically justified closure assumptions and relaxation terms.
A preliminary one-dimensional numerical implementation of the model is shown to give satisfactory results in a test case involving a high-velocity collision of two plasma flows.
Paths to extend that model to three dimensions as needed for an actual hohlraum geometry are briefly discussed.
\vskip 2em
\begin{multicols}{2}
\section{Introduction}\label{sec:intro}
In indirectly driven Inertial Confinement Fusion (ICF) \cite{LIN981,LIN041,ATZ042}, a capsule containing the thermonuclear fuel is placed inside a high-$Z$ material case (``hohlraum'') heat\-ed by powerful laser beams.
In addition to generating thermal radiation driving the capsule, the laser beams create a plasma filling the hohlraum.
In some designs, e.g. the so-called Near Vacuum Hohlraum (NVH)\cite{BER15H}, that plasma is so tenuous and hot that the mean-free path for Coulomb collisions among the plasma ions is no more negligible.
As a result, the density calculated by standard hydrodynamics codes in regions where plasma flows collide at high velocity inside the hohlraum can be in error, leading to spurious deviations of the heating laser beams, which in turn can alter the calculated symmetry of the capsule implosion.

This can occur in various places inside the hohlraum, e.g. where the expanding plasmas from the hohlraum case and the capsule ablator meet and possibly interpenetrate, or on the hohlraum axis where the case plas\-ma collides onto itself (see Fig.~1).
Although those plas\-ma collision phenomena are real and have been actually observed in experiments \cite{GLE992,DAT019,BAC037}, there is growing concern that they are not properly accounted for in large hydrodynamics codes.

This erroneous behaviour has been mitigated in numerical simulations \cite{BER15H,HIG216} by artificially increasing the laser beam frequency above the critical density prevailing in the spurious density ridges arising from plasma collision, thus bringing the beams back onto their expected propagation path.
However, this trick has other undesirable consequences (e.g., on laser absorption by the hohl\-raum case and conversion into thermal X-rays), and cannot be considered a satisfactory solution of the problem.

To specifically investigate the effect of increased collisional mean-free paths in plasma collisions, dedicated experiments have been designed, aiming at reproducing that interaction in a more controllable way \cite{CHE979,WAN979,FAR999,REN13F,FAL155,RIN18D}.
To specifically study the interaction of the ablator and hohlraum case materials, an experimental setup has recently been used \cite{LEP208}, which involves the ablative expansion of a carbon plasma and a gold plasma facing each other, initially separated by an adjustable amount of helium gas.
It was indeed found that, for parameters comparable to those inside an ICF hohlraum, the ablator and hohlraum plasmas did interpenetrate more or less, depending on the amount of He gas initially present, instead of only stagnating against each other as predicted by standard hydrodynamics.
This behaviour can be qualitatively reproduced using our multi-fluid numerical code \textsc{multif} \cite{CHE979}.

However, multifluid models cannot be used to simulate the collision of a single plasma onto itself in a convergent geometry, as is the case near the hohlraum axis (see Fig. 1), because in such a situation there is no way to split the distribution function of the plasma ions in two (or more) well-separated components in velocity space \cite{RAM94C}.
A method for taking into account the possibly large deviations of the distribution from the equilibrium Maxwellian is thus needed, independently from the treatment of genuinely multifluid situations.
This is also what happens in the fuel gas contained in strong\-ly kinetic exploding-pusher ICF targets \cite{ROS14S,SIO19F}, %,ROS15O,
where two-component, multifluid-like features develop in the course of the implosion, with sizeable consequences on the implosion metrics (neutron yield, ion temperature, etc...) \cite{LAR158,LAR183}.

There is thus a need for a capability of numerically simulating two different kinds of ion-kinetic effects:
i)~the interpenetration of different flows and/or species which obviously requires a multifluid treatment, and
ii)~the possibly strong \rule{2em}{0pt}

{\setlength{\parindent}{0cm}
\parbox{\linewidth}{\small
\rule{0pt}{1em}
\begin{center}
\includegraphics[scale=0.65]{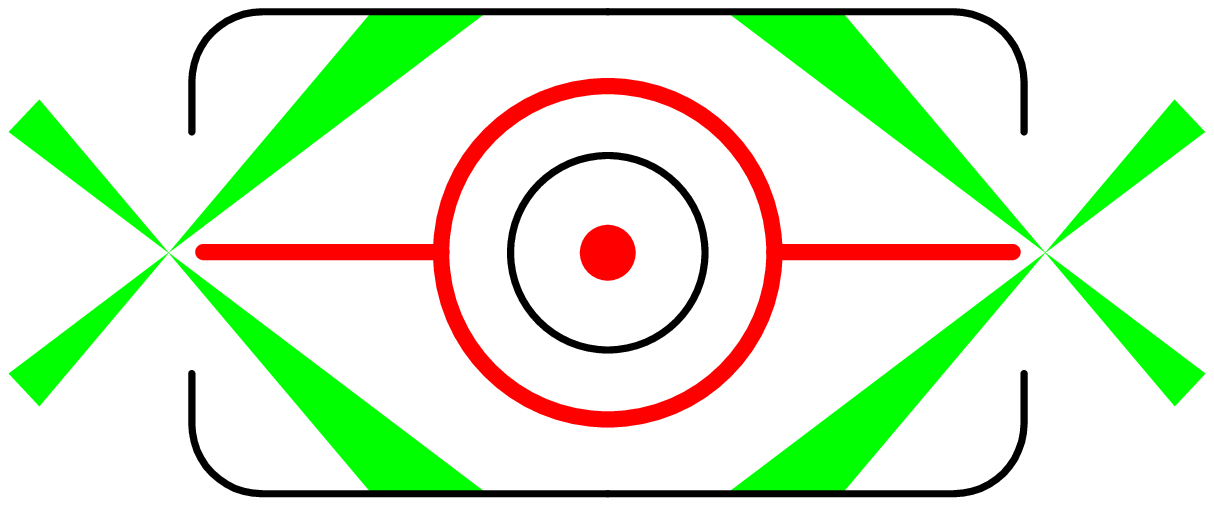}
\end{center}

\textbf{Fig. 1.}\rule{1em}{0pt}Schematic of the plasma collision regions inside an ICF hohlraum. Collision/interpenetration can occur on the case axis, between the case and capsule ablator, or in the fuel contained in the capsule (regions marked in red). The laser beams heating the hohlraum interior are represented in green.}}
deviations from thermodynamic equilibrium of a single species, which requires to go beyond the standard treatment based on Euler or Navier-Stokes equations.

Several tools have been developed, in various fields of application, to tentatively fulfill those requirements.At the most fundamental level, kinetic codes directly solve the Boltzmann equation.
Such tools have been developed, following two different strategies: on one hand, ``deterministic'' codes solving the Vlasov-Fokker-Planck equation for the
discretized ion distribution function \cite{LAR183,LAR931,TAI214}, and on the other hand, particle-in-cell codes statistically sampling the distribution ``\`a la Monte Carlo'' \cite{RAM94C,RAM952,ARB15B,BIR216}.
Tho\-se codes provide reference solutions and physical insight in some simplified or academic situations \cite{LAR931,RAM952,LAR03A,BEL149,PEI14A,SAG147}, but need too much computer time to be used for the routine simulation of hohlraums.

The multi-species interpenetration problem has been traditionally treated by multifluid models in a 5-mo\-ment, Euler equation formalism \cite{CHE979,BER917,RAM942}.
That approach has received a renewed interest recently \cite{GHO195,MAR212}.
A multifluid capability has also recently been implemented \cite{HIG216} into the standard radiative hydrodynamics code \textsc{lasnex} \cite{ZIM752}.
Other approaches include hybrid, single-ave\-rage-fluid/multiple velocity models \cite{CHA122}.

On the other hand, the departure from local equilibrium of a single species has been tackled through extended hydrodynamics models, beyond the 5-moment frame, where additional moments of the velocity distribution are treated as independent, dynamical variables.
This approach can be traced back to the classic work of Grad \cite{GRA491}, which handles 13 moments, namely: the density, and the components of the velocity vector, pressure tensor and heat flux vector.

For a single particle species, it is generally held for true that the hierarchy of moment equations is equivalent to an expansion of the Boltzmann equation in terms of a collisionality parameter such as the Knudsen number (see, e.g., the discussion by Levermore \cite{LEV961}).
It is then natural to expect that the behaviour of the moment equations will tend to match that of moments of the kinetic equation when an increasing number of moments are considered.
The detailed way in which this convergence proceeds is very nicely illustrated in the paper by Au \textit{et al} \cite{AUU013}, where it is shown that the increasing number of hyperbolic waves obtained in the moment equations, propagating with an increasingly larger set of characteristic velocities, yield increasingly refined piecewise approximations of the kinetic moment profiles which get closer and closer to them as the number of moments gets larger.
This is the physical mechanism which garantees that even, say, an interpenetration situation can be described satisfactorily by a sufficiently large set of moment equations.
In a different physical context (that of multiphase flow), this is strikingly illustrated by the results of Ref.~\cite{FOR192}.

Now in a practical situation as the hohlraum problem, the question is: how many moments need to be taken into account (and exactly how should they be chosen) for the main mechanisms at play to be reasonably rendered.
In this work it is claimed (and tentatively demonstrated in a simple one-dimensional case) that moments of order three already give a useful description of important hydrodynamical consequences of the interpenetration process, in particular that unphysical density ridges arising in the frame of the classical Euler equations are smoothed out.
This can be understood qualitatively from the moments of the velocity distribution for interpenetrating (or interpenetrating-component) plasmas: if the two components have identical densities, then the distribution will be symmetrical around the common bulk velocity, and a 10-moment description (as in Ref.~\cite{BAR107}) will be sufficient.
But if the densities are different, the resulting asymmetry will translate into a large heat flux, which makes it necessary to take into account the third-order moments.

Two more ingredients are needed to complete the model, namely
i)~an assumption about the underlying velocity distribution providing an estimate of the ``missing moments'' to close the system of equations obtained, and
ii)~expressions for the r.h.s. relaxation terms which act to bring the system back to local thermodynamic equilibrium.
In the case of Grad's 13-moment model, the closure relies on an expansion of the underlying distribution on a basis of Hermite orthogonal polynomials, and the relaxation terms are calculated for various types of collision differential cross-section, e.g., the Maxwellian inverse-fifth-power molecule or elastically colliding rigid spheres.
In the present case of Coulomb collisions, of course this should be replaced with the corresponding forward-peaked cross-section, asymptotically vanishing for large relative velocity, leading to the Fokker-Planck form of the relaxation terms \cite{ROS573}, and this has important consequences in plasma-collision situations.

Another point which must not be overlooked is the hyperbolicity of the system of conservation equations thus obtained.
From a loose, physicist's point of view, this is interpreted as checking whether short-wavelength linear modes of the system can become unstable in some regions of the parameter space, which can obviously lead to a breakdown of the simulation.
As it turns out \cite{TOR001}, Grad's classic 13-moment system is hyperbolic only in a restricted domain of the parameter space defined by the pressure anisotropy and the heat flux, and as a result, should only be used with due care.

Nonetheless, $N$-moment methods have been implemented in many fields of computational hydrodynamics.
To name a few:
\begin{itemize}
\item numerous attempts in interplanetary or interstellar physics (e.g., the solar wind problem, or the dynamics of gravitational systems), dating as far back as \cite{LAR701} or \cite{CUP813}; for a review, see \cite{ECH113};
\item magnetosphere physics \cite{NGG214}, with a 10-moment, ``pa\-rabolic'' closure (through a Fourier description of the heat flux $\vec{q}=-\kappa\vec{\nabla}T$);
\item multiphase flows and aerosols \cite{FOR192}: a 14-moment, entropic closure with interpolation;
\item rarefied gases \cite{BOH202}: $N$-moment with various values of $N$, quadrature closure (``QMOM'');
\item a general problem with a specific closure \cite{TOR10A}: 13-moment with a Pearson-Type-IV underlying distribution;
\item magneto-hydrodynamics (MHD) and ``magnetized liner inertial fusion'' \cite{HAM212}: an 8-moment (density, velocity vector, isotropic pressure and heat flux vector), polynomial closure.
\end{itemize}
This paper will present a new 10-moment (density, velocity vector, unit vector of azimuthal symmetry, parallel and transverse pressure, parallel and transverse heat flux) model with a specific closure appropriate for plasma collision situations.
Once again, there are indeed two kinds of physical effects which need to be taken into account: on one hand, the non-equilibrium features occurring in the velocity distribution function of each ion species, and on the other hand the fact that different species can behave differently, leading to interpenetration/separation effects.
In this work only the first one is investigated, but obviously, to get a complete treatment of kinetic effects, both aspects must be treated.
This will be done by multifluid codes (such as described in Refs.~\cite{CHE979,BER917,RAM942,GHO195,MAR212}), in which each species will be described by an extended hydrodynamics, N-moment formalism.

Actually, there have already been some attempts at including both multifluid and $N$-moment features in a single numerical model:
\begin{itemize}
\item our own 10-moment, vanishing-heat-flux multifluid code \textsc{multif} \cite{CHE979} (actually, since it is only one-dimen\-sional, that model should rather be called 4-moment) with a specific, anisotropy-aware, relaxation term \cite{BAR107};
\item a 13-moment multifluid code for MHD \cite{MIL162} with Pearson-IV closure and ``Bhatnagar-Gross-Krook'' (BGK) \cite{BHA543} collisional relaxation (i.e., not taking into account the dependence of relaxation rates on aniso\-tropy).
\end{itemize}
It should be noticed that interpenetration features (such as a double-humped shape) can arise in the velocity distribution of a single ion species due to a strong collision with another plasma occurring nearby, even if there is no sizeable mixing between the two: this is indeed what is found in kinetic simulations of the inner gas in strongly kinetic, ``exploding-pusher-like'' ICF capsule implosions (see Figs.~6 and 7 of Ref.~\cite{LAR158}).
In some of those shots, although the fuel/pusher interaction is sufficiently collisional that they interpenetrate only marginally, the hot tenuous plasma comprising the fuel is itself highly kinetic, so that the ``snowplough'' effect from the pusher generates a two-beam structure in the velocity distribution of the fuel.
It is only for still more strongly driven implosions that the pusher and fuel plasmas start to interpenetrate, which needs to be treated accordingly \cite{LAR183}.
Such a snowplough effect can also be expected in the tenuous gas of low-fill-density hohlraums \cite{BER15H}, and thus needs to be accounted for in the extended-hydrodynamics formalism for each species.
This is what guided our choice of an appropriate moment set and closure distribution.
We took care that this choice can also treat other non-equilibrium features, such as the negative pressure anisotropy which can arise in an expanding plasma flow or due to collisional relaxation in the interpenetration with another plasma.

The rest of this paper will be organized as follows.
Section \ref{sec:kintohydro} will recall the basic equations and review the process of going from the kinetic description to the $N$-moment model in the present case of ICF hohlraum physics.
In particular, the specific differential cross-sec\-tion for Coulomb collisions in a plasma will be shown to allow flow interpenetration.
Section \ref{sec:axisclos} will describe the specific closure introduced in this work, taking into account the main features expected from the underlying distribution in a context where plasma collisions are expected.
Section \ref{sec:impl1D} will apply the general formalism derived to the case of a plane one-dimensional geometry, and investigate the hyperbolicity of that reduced implementation.
Section \ref{sec:numres} will present first results obtained with that new model in a test case involving the collision of two plasma slabs at high velocity.
Finally Sect. \ref{sec:conclu} will summarize the results and discuss the work needed to implement the present model in a more general, three-dimensional geometry.

\section{From kinetic theory to hydrodynamics: moments, closure, relaxation}
\label{sec:kintohydro}

The starting point is the Boltzmann equation governing the evolution in time $t$ of the velocity distribution $f(\vec{x},\vec{c})$ for a given species of ions of mass $m$ in configuration ($\vec{x}$) and velocity ($\vec{c}$) space, with non-collisional, advection terms on the l.h.s., and collisional relaxation terms on the r.h.s.:
\begin{equation}
\frac{\partial f}{\partial t} + c_i\frac{\partial f}{\partial x_i} + \frac{F_i}{m}\frac{\partial f}{\partial c_i} = \mathcal{C}(f) \label{eq:Bol}
\end{equation}
The vector $\vec{F}$ is an external force acting on the ions, e.g. the force $Ze\vec{E}$ exerted on ions of charge $Ze$ by the ambipolar electric field $\vec{E}$ in a non-homogeneous plasma.
Here and in the following, summation over repeated indices is assumed.
That description includes the particle translation degrees of freedom, but not their possible internal degrees of freedom.
Going beyond that is a problem in itself, see on that point Refs. \cite{AUU013,MCC681,JOU872}.
Second-order correlations and dense-plasma effects \cite{STA212} will be neglected, leading to a perfect gas equation of state.
However this makes sense since the kinetic effects we study occur in moderately collisional situations involving tenuous plasmas.

From the kinetic equation we want to derive evolution equations for a small (or at least not too large) number of macroscopic quantities, expressed from velocity moments of the distribution function. There are two main types of procedure for doing so (see, e.g., the very pedagogical discussion at the beginning of Refs. \cite{TOR10A,LEV961}).
The first one is the ``Chapman-Enskog expansion'' in the vicinity of local equilibrium in the limit of strong collisions, in which only the first non-trivial order in terms of the collision time is retained, which leads to the Navier-Stokes equation with Fourier heat conduction.
Higher orders lead to various issues \cite{STR04A}, and as a consequence the resulting equations are not routinely used in practical simulations.
This procedure rests on the assumption that the characteristic time scales of the system are much longer than the collisional relaxation time, which, at lowest order, leads to a quasi-stationary equilibrium where the time derivative is dropped from Eq. (\ref{eq:Bol}).
That equilibrium then evolves adiabatically according to the time derivatives kept at the next expansion order.
The procedure breaks down when the relaxation processes are not strong enough to enforce that low-order equilibrium.
The second procedure is the expansion of the distribution into moments of increasing powers of the velocity, which needs to be stopped and ``closed'' through various types of hypotheses about the underlying distribution; these yield expressions for both the missing higher-order moments on the l.h.s., and the collisional relaxation terms on the r.h.s.

The systems obtained through moment methods can themselves be closed by a Chapman-Enskog expansion, which leads to the so-called ``regularized'' $N$-moment methods \cite{STR032,TIM156}.
An early example of this is given by Candler \textit{et al} \cite{CAN944} who describe a shock front by keeping the anisotropy of the pressure tensor in order 0 of a Chapman-Enskog expansion, while keeping collisional relaxation terms for the pressure anisotropy in the r.h.s., leading to a hybrid method.
This was further developed by Xu and Josyula \cite{XUU05B} using a BGK kinetic numerical scheme with various technical procedures to recover the shock width found in the completely kinetic Direct Simulation Monte Carlo (DSMC) simulation of Ref. \cite{CAN944}.
In the same spirit, see also Holway \cite{HOL661}.

In the rest of this paper, the needed tensorial moments of the velocity distribution will be defined as follows:
\begin{align*}
    \rho &= m\int f(\vec{c})d^3c \\
j_i = \rho v_i &= m\int c_if(\vec{c})d^3c \\
P_{ij}   &= m\int (c_i-v_i)(c_j-v_j)f(\vec{c})d^3c \\
Q_{ijk}  &= m\int (c_i-v_i)(c_j-v_j)(c_k-v_k)f(\vec{c})d^3c
\end{align*}
\begin{align*}
R_{ijkl} &= m\int (c_i-v_i)(c_j-v_j)\times \\
         & \hskip 5em (c_k-v_k)(c_l-v_l)f(\vec{c})d^3c
\end{align*}
From these, the scalar pressure $P$ and heat flux vector $\vec{q}$ are defined as
\begin{align*}
P   &= \frac{1}{3}P_{ii} \\
q_k &= \frac{1}{2}Q_{iik}
\end{align*}

\subsection{Moments up to order 3}

Moments of the kinetic equation (\ref{eq:Bol}) are taken, leading to evolution equations for integrals of the monomials $1$, $\vec{c}$, $c_ic_j$ and $c_ic_jc_k$, taking into account the tensor definitions given above.
In this work the expansion will include moments of order 3 (leading to evolution equations for $Q_{ijk}$), improving over a previous description involving moments of order 2 only \cite{BAR107} which was restricted to symmetric situations where the heat flux was expected to vanish, such as the interpenetration of identical plasmas.
It will be checked that more general situations can be satisfactorily described without having to use still higher moments.
The present investigation will be restricted to a velocity-independent external force $F_i$, and thus to the case of a vanishing magnetic field.
The following system is obtained:
\begin{align}
& \frac{d\rho}{dt} + \rho\frac{\partial v_i}{\partial x_i} = 0 \label{eq:drhodt} \\
& \frac{dv_i}{dt} + \frac{1}{\rho}\frac{\partial P_{ij}}{\partial x_j} = \frac{1}{m}F_i +\left(\frac{\partial v_{i}}{\partial t}\right)_c \label{eq:dvdt} \\
& \frac{dP_{ij}}{dt} + P_{ij}\frac{\partial v_k}{\partial x_k} +2P_{\underline{i}k}\frac{\partial v_{\underline{j}}}{\partial x_k} +\frac{\partial Q_{ijk}}{\partial x_k} = \left(\frac{\partial P_{ij}}{\partial t}\right)_c \label{eqdPdt} \\
& \frac{dQ_{ijk}}{dt} + Q_{ijk}\frac{\partial v_l}{\partial x_l} + 3Q_{\underline{ij}l}\frac{\partial v_{\underline{k}}}{\partial x_l} -\frac{3}{\rho}P_{\underline{ij}}\frac{\partial P_{\underline{k}l}}{\partial x_l} +\frac{\partial R_{ijkl}}{\partial x_l} \nonumber \\
 & \hskip 15em = \left(\frac{\partial Q_{ijk}}{\partial t}\right)_c \label{eq:dQdt}
\end{align}
where terms with subscript $c$ on the r.h.s. stand for the integrals of the corresponding velocity monomials against the collision kernel $\mathcal{C}(f)$.
In those expressions, terms with underscored indices stand for their symmetrized form with respect to permutations of the given indices (i.e., the sum of all permuted terms divided by the number of permutations), which for tensors $\mathbf{T}$ of order 2 and 3, reads:
\begin{align*}
& T_{\underline{ij}} = \frac{1}{2}(T_{ij}+T_{ji}) \\
& T_{\underline{ijk}} = \frac{1}{6}(T_{ijk}+T_{jik}+T_{ikj}+T_{jki}+T_{kij}+T_{kji})
\end{align*}
To display more clearly the effect of source terms on the r.h.s. on the evolution of hydrodynamic quantities, the above system is written in terms of the convective derivative ($\frac{d.}{dt}=\frac{\partial.}{\partial t}+v_i\frac{\partial.}{\partial x_i}$) of variables of successive orders.
This form is no longer conservative, but is interesting because it is quasi-linear with respect to the gradients of advected quantities:
\[
\frac{d\mathbf{u}}{dt} + \mathbf{A}_l\frac{\partial\mathbf{u}}{\partial x_l} = ...
\]
where $\mathbf{u}=(\rho,v_i,P_{ij},Q_{ijk})^t$ is the vector of advected quantities. This form can be useful to investigate the linear stability of the system (``hyperbolicity'', see Section \ref{sec:hyperb}).

It can be checked that (leaving aside the external force terms) Eqs. (5.17) of Grad \cite{GRA491} are recovered, in which the moment of order 4 was replaced by its approximation (Eq. (5.16)) from the distribution expansion to order 3 in Hermite polynomials, namely:
\[
R_{ijkl} = \frac{P}{\rho}\left(6P_{\underline{ij}}\delta_{\underline{kl}}-3P\delta_{\underline{ij}}\delta_{\underline{kl}}\right)
\]
but moments of order 3 have not yet been replaced by the ``Grad's 13-moment'' approximation (Eq. (5.9)) which reads:
\[
Q_{ijk} = \frac{1}{5}(3\delta_{\underline{ij}}Q_{\underline{k}ll}) = \frac{2}{5}(3\delta_{\underline{ij}}q_{\underline{k}})
\]

\subsection{Closure and relaxation in a plasma}
\label{sec:closplas}

To proceed without having to keep too many moments, additional closure assumptions must be made about the components of the moment of order 3 and the tensor of order 4 $R_{ijkl}$.
Also, the collision terms on the r.h.s. of the equations must be given actual values.
In a plasma where the collision process arises from the Coulomb electric interaction, those terms are velocity integrals of the Fokker-Planck collision terms \cite{ROS573} (see Appendix \ref{app:FP}), leading to characteristic collision times $\tau_c\propto\Delta v^3$ where $\Delta v$ is the relative velocity of the colliding particles.

Due to that scaling, collisions between ions and electrons, for comparable temperatures $T_i$ and $T_e$, are much weaker than collisions between ions.
This work focusses on ion-ion collisions, and their role in hydrodynamics.
Electron-ion collisions are treated using the usual relaxation coefficients (see, e.g., \cite{BRA65A,DEC981}), which are not different whether the ion distribution is in thermodynamic equilibrium or not, provided the global values of the ion moments (bulk velocity, pressure, and so on) are used in the formulas.
This is true as long as the electron thermal velocity remains much larger than the ion velocities, which is assumed to be the case in ICF plasmas.

Because of the $\Delta v^3$ scaling of collision times, a situation such as illustrated in Fig. 2 can arise when two plasma flows collide at high velocity.
If the relative velocity $v$ between the two flows is large, $v \gg \left(\frac{k_BT_\alpha}{m}\right)^{1/2}$ where $T_\alpha=P_{ii}^{(\alpha)}/(3k_Bn_\alpha)$ is the temperature of flow $\alpha$, $\alpha=1$ or 2, and $n_\alpha$ is the particle density of component $\alpha$, the various collision mechanisms proceed over different characteristic times, namely, focusing on flow~1:
\begin{itemize}
\item
a self-thermalization time
(see Eq. (\ref{eq:tauab}) and\linebreak Fig.~2a):
\begin{equation}
\tau_{D\parallel} \sim \frac{1}{n_1} \left(\frac{k_BT_1}{m}\right)^{3/2} \label{eq:therm}
\end{equation}
\item
an isotropisation (or angle diffusion) time
(see Eq. (\ref{eq:tauDab}) and Fig.~2b):
\begin{equation}
\tau_{D\bot} \sim \frac{1}{n_2} \frac{k_BT_1}{m}\left(v^2+\frac{\pi}{2}\frac{k_BT_2}{m}\right)^{1/2} \label{eq:isotr}
\end{equation}
\item
a slowing-down time
(see Eq. (\ref{eq:tauRab}) and Fig.~2c):
\begin{equation}
\tau_R \sim \frac{1}{n_2}\left(v^2+\left(\frac{9\pi}{2}\right)^{1/3}\frac{k_BT_2}{m}\right)^{3/2} \label{eq:ralen}
\end{equation}
\end{itemize}
For $T_1\approx T_2$ and $n_1\approx n_2$, we get: $\tau_{D\parallel}\ll\tau_{D\bot}\ll\tau_R$ and the relaxation to the global Maxwellian occurs only over the longest time ($\tau_R$).
Over shorter durations $\Delta t$ such that $\tau_{D\parallel}<\Delta t<\tau_R$, the distribution is in a ``metastable'' state where the individual components are close to thermodynamic equilibrium, while slowly drifting (in velocity space) towards each other, so that the collisional relaxation of that system is essentially a slowing-down process.
This is why interpenetration is expected to occur in plasmas, which justifies the treatment of such situations by multifluid codes.
On that basis, we will use a specific closure involving a two-component distribution, described in the following section, together with heuristic expressions for the needed relaxation rates.

The idea of using two-component velocity distributions is not new \cite{MOT511}, and has been used to describe the structure of a stationary shock wave in a plasma \cite{ABE852}.
However, in the latter case, that description can lead to erroneous conclusions, e.g. regarding the excitation of plasma waves, because the actual distribution in the shock front, as found in kinetic computations \cite{CAS91A,VID954}, is hardly double-humped.
On the contrary, in the present case of the dynamic non-stationary interpenetration of plasma streams, as demonstrated above there is a time-scale range over which a metastable two-stream structure can persist, and needs to be treated in its own right.

\section{A specific closure for plasma collision situations}
\label{sec:axisclos}
Having in mind the type of distribution illustrated in Fig. 2, we investigate a 10-moment closure resting on an underlying distribution with two axisymmetric components (see Fig. 3):
{\setlength{\parindent}{0cm}
\parbox{\linewidth}{\small
\rule{0pt}{2em}
\begin{center}
\includegraphics[scale=0.25]{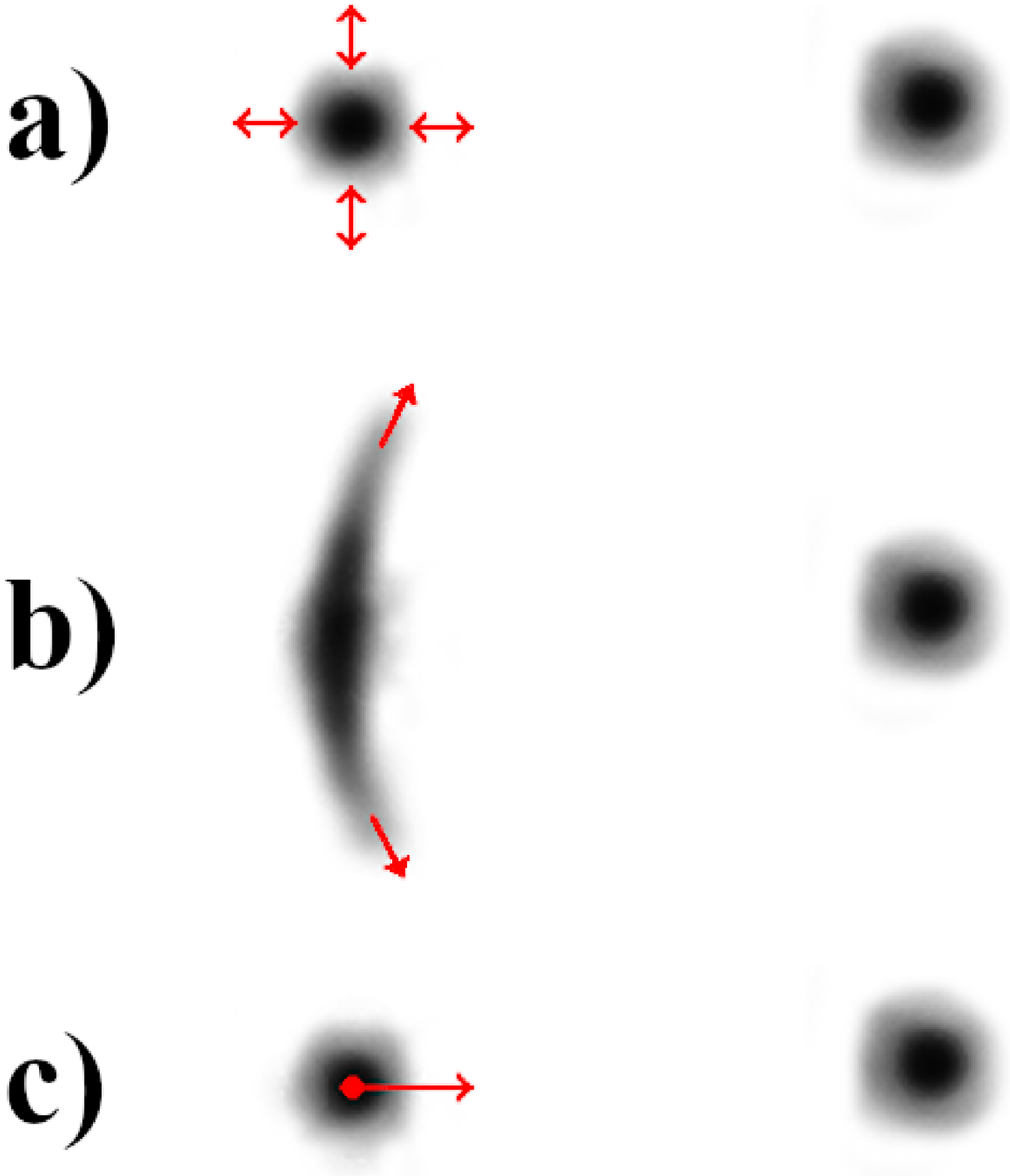}
\end{center}

\textbf{Fig. 2.}\rule{1em}{0pt}Relaxation effects on the velocity distribution from the three terms in the Fokker-Planck collision operator, when the distribution consists of two interpenetrating components: a) thermalization: collisions among particles of the single component 1 (on the left); b) angle diffusion of component 1 from collisions on component 2 (on the right); c) slowing-down of component 1 from collisions on component 2.
}}

{\setlength{\parindent}{0cm}
\parbox{\linewidth}{\small
%\rule{0pt}{2em}
\begin{center}
\includegraphics[scale=0.65]{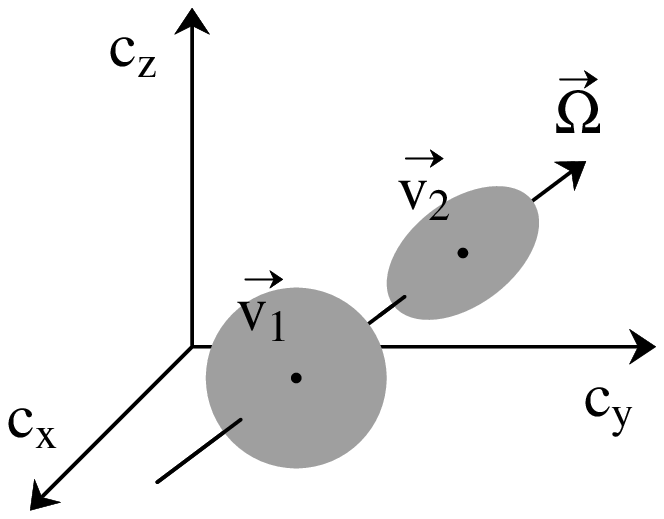}
\end{center}

\textbf{Fig. 3.}\rule{1em}{0pt}Schematic of the interpenetrating distribution function used to close the moment hierarchy.

\rule{0pt}{2em}
}}
\[
f(\vec{c}) = \frac{\rho_1}{m}f_\parallel^{(1)}(\vec{c})f_\bot^{(1)}(\vec{c}) + \frac{\rho_2}{m}f_\parallel^{(2)}(\vec{c})f_\bot^{(2)}(\vec{c})
\]
which is azimuthally symmetric around the vector
\[
\vec{\Omega} = \frac{\vec{v}_2-\vec{v}_1}{|\vec{v}_2-\vec{v}_1|}
\]
With that symmetry hypothesis, each component $n=1$ or 2 has an anisotropic pressure tensor
\[
P_{ij}^{(n)} = \rho_n\frac{k_BT_{\parallel n}}{m}\Omega_i\Omega_j + \rho_n\frac{k_BT_{\bot n}}{m}(\delta_{ij}-\Omega_i\Omega_j)
\]
The macroscopic parameters (velocity moments) for the resulting distribution read (see Appendix \ref{app:moms}):
\begin{align}
\rho &= \rho_1+\rho_2 \nonumber \\
\rho\vec{v} &= \rho_1\vec{v}_1+\rho_2\vec{v}_2 \nonumber \\
P_{ij} &= \Ppa \Omega_i\Omega_j + \Ppe (\delta_{ij}-\Omega_i\Omega_j) \label{eq:Pijaxis} \\
Q_{ijk} &= \Qpa \Omega_i\Omega_j\Omega_k + \Qpe [\Omega_i(\delta_{jk}-\Omega_j\Omega_k) \nonumber \\
         &\hskip 3em +\Omega_j(\delta_{ik}-\Omega_i\Omega_k)+\Omega_k(\delta_{ij}-\Omega_i\Omega_j)] \label{eq:Qijkaxis}
\end{align}
where
\begin{align*}
\Ppa &= \rho_1\frac{k_BT_{\parallel 1}}{m}+\rho_2\frac{k_BT_{\parallel 2}}{m} +\frac{\rho_1\rho_2}{\rho_1+\rho_2}|\vec{v}_2-\vec{v}_1|^2 \\
\Ppe &= \rho_1\frac{k_BT_{\bot 1}}{m}+\rho_2\frac{k_BT_{\bot 2}}{m} \\
\Qpa &= \frac{\rho_1\rho_2}{\rho_1+\rho_2}|\vec{v}_2-\vec{v}_1|\left[3\left(\frac{k_BT_{\parallel 2}}{m}-\frac{k_BT_{\parallel 1}}{m}\right)\right. \\
        & \hskip 10em \left. +\frac{\rho_1-\rho_2}{\rho_1+\rho_2}|\vec{v}_2-\vec{v}_1|^2\right] \\
\Qpe &= \frac{\rho_1\rho_2}{\rho_1+\rho_2}|\vec{v}_2-\vec{v}_1|\left(\frac{k_BT_{\bot 2}}{m}-\frac{k_BT_{\bot 1}}{m}\right)
\end{align*}
In passing, the heat flux vector of the resulting distribution  is
\[
\vec{q} = \left(\frac{1}{2}\Qpa+\Qpe\right)\vec{\Omega}
\]
The resulting distribution is thus defined by ten independent variables: $\rho$, the three components of $\vec{v}$, $\Ppa$, $\Ppe$, $\Qpa$, $\Qpe$ and the two Euler angles defining the unit vector $\vec{\Omega}$.
On the other hand, each component of the distribution is defined by six variables:
$\rho_n$, $\vec{v}_n$, $P_{\parallel n}$ and $P_{\bot n}$.
Two additional constraints are thus needed to unambiguously define the closure distribution from the ten macroscopic variables given.
Since reproducing any possible values of $\Ppe$ and $\Qpe$ needs unconstrained values of $T_{\bot 1}$ and $T_{\bot 2}$, the only possibility is to link $T_{\parallel 1}$ and $T_{\parallel 2}$ to the bulk velocity difference $|\vec{v}_2-\vec{v}_1|$.
The simplest choice is to let $T_{\parallel 1}=T_{\parallel 2}=0$.
A less singular choice is a ``double-waterbag'' distribution such as illustrated on Fig. 4:
defining $c_\parallel = (\vec{c}-\vec{v})\cdot\vec{\Omega}$ (velocity along the symmetry
{\setlength{\parindent}{0cm}
\parbox{\linewidth}{\small
%\rule{0pt}{2em}
\begin{center}
\includegraphics[scale=0.75]{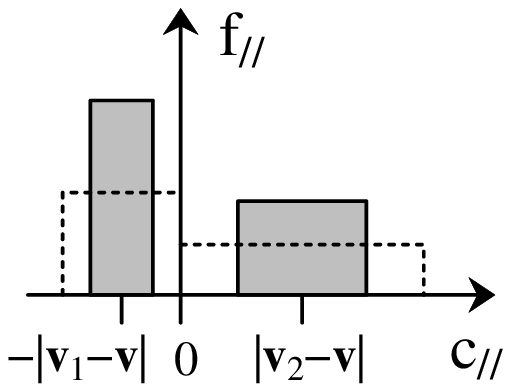}
\end{center}

\textbf{Fig. 4.}\rule{1em}{0pt}Schematic of the parallel velocity distribution function in the ``double-waterbag'' closure; $c_\parallel$ is the velocity along the symmetry axis: $c_\parallel = (\vec{c}-\vec{v})\cdot\vec{\Omega}$ . The relative width of the two flat-top components with respect to their velocity difference is expressed by a parameter $\epsilon$ which can be chosen between 0 (for vanishing widths) and 1 (for maximum width as displayed by the dashed lines in the figure).

\rule{0pt}{2em}
}}
axis), the parallel velocity distribution functions (as defined in Eq.~(\ref{eq:Cfparn})) for the components are taken in the form
\[
f_\parallel^{(1)}(\vec{c}) = \frac{1}{2\epsilon|\vec{v}-\vec{v}_1|}
\]for $-(1+\epsilon)|\vec{v}-\vec{v}_1| < c_\parallel < -(1-\epsilon)|\vec{v}-\vec{v}_1|$, otherwise 0; and
\[
f_\parallel^{(2)}(\vec{c}) = \frac{1}{2\epsilon|\vec{v}-\vec{v}_2|}
\]
for $(1-\epsilon)|\vec{v}-\vec{v}_2| < c_\parallel < (1+\epsilon)|\vec{v}-\vec{v}_2|$, otherwise 0.
This form goes over to a set of two delta-functions centered on $|\vec{v}-\vec{v}_1|$ and $|\vec{v}-\vec{v}_2|$ when $\epsilon\to 0$, and to the dashed curves on Fig. 4 when $\epsilon\to 1$.
The choice of a compact support for the parallel distribution is handy for the numerical implementation of the closure \cite{PER92A}.
Among other things, it gives a straightforward derivation of the CFL condition for the scheme (however, an equivalent CFL condition can still be defined for non-compact support, see \cite{EST962}).
All physical values of $\rho$, $\vec{v}$, $\Ppa$, $\Ppe$, $\Qpa$, $\Qpe$ are realizable, i.e., can be unambiguously translated into a set of parameters for the underlying component distributions, as follows.
If we define
\[
\theta = \mathrm{argsh}\left(\frac{(3+\epsilon^2)^{3/2}}{2(1+\epsilon^2)}\frac{\rho^{1/2}\Qpa}{\left(3\Ppa\right)^{3/2}}\right) % q = shθ
\]
then
\[
\rho_1 = \rho\frac{1+\tgh\theta}{2} \quad,\quad \rho_2 = \rho\frac{1-\tgh\theta}{2}
\]
and the component parallel distributions take on the following values inside their respective supports:
\[
f_\parallel^{(1)} = \frac{(3+\epsilon^2)^{1/2}}{2\epsilon}\left(\frac{\rho}{3\Ppa}\right)^{1/2}\mathrm{e}^\theta 
\]
and
\[
f_\parallel^{(2)} = \frac{(3+\epsilon^2)^{1/2}}{2\epsilon}\left(\frac{\rho}{3\Ppa}\right)^{1/2}\mathrm{e}^{-\theta}
\]
The remaining component parameters follow:
\begin{align*}
& \vec{v}_1 = \vec{v}-\left(\frac{3\Ppa}{(3+\epsilon^2)\rho}\right)^{1/2}\mathrm{e}^{-\theta}\vec{\Omega} \\
& \vec{v}_2 = \vec{v}+\left(\frac{3\Ppa}{(3+\epsilon^2)\rho}\right)^{1/2}\mathrm{e}^{\theta}\vec{\Omega} \\
& \frac{k_BT_{\parallel 1}}{m} = \frac{\epsilon^2}{3+\epsilon^2}\frac{\Ppa}{\rho}\mathrm{e}^{-2\theta} \\
& \frac{k_BT_{\parallel 2}}{m} = \frac{\epsilon^2}{3+\epsilon^2}\frac{\Ppa}{\rho}\mathrm{e}^{2\theta} \\
& \frac{k_BT_{\bot 1}}{m} = \frac{\Ppe}{\rho} - (1+\epsilon^2)^{1/2} \frac{\Qpe}{(3\rho\Ppa)^{1/2}} \mathrm{e}^{-\theta} \\
& \frac{k_BT_{\bot 2}}{m} = \frac{\Ppe}{\rho} + (1+\epsilon^2)^{1/2} \frac{\Qpe}{(3\rho\Ppa)^{1/2}} \mathrm{e}^{\theta}
\end{align*}
The parameters of the fourth-order tensorial moment (see Appendix \ref{app:moms})
\begin{multline}
R_{ijkl} = R_{\parallel\parallel} \Omega_i\Omega_j\Omega_k\Omega_l
  + R_{\parallel\bot}[\Omega_i\Omega_j(\delta_{kl}-\Omega_k\Omega_l) + ...\ ] \\
      +R_{\bot\bot} [(\delta_{ij}-\Omega_i\Omega_j)(\delta_{kl}-\Omega_k\Omega_l) +...\ ] \label{eq:Rijkmaxis}
\end{multline}
which closes the system are given by Eqs. (\ref{eq:Rpapa2})-(\ref{eq:Rpepe}).
Those needed for the one-dimen\-sional implementation investigated in Sect. \ref{sec:impl1D} are
\begin{align}
\Rpapa &= p_\parallel\frac{\Ppa^2}{\rho} + q_\parallel\frac{\Qpa^2}{\Ppa} \label{eq:Rpapa} \\
\Rpape &= p_\bot\frac{\Ppa\Ppe}{\rho} + q_\bot\frac{\Qpa\Qpe}{\Ppa} \label{eq:Rpape}
\end{align}
where
\begin{align}
p_\parallel &= \frac{1+2\epsilon^2+\frac{1}{5}\epsilon^4}{\left(1+\frac{\epsilon^2}{3}\right)^2} \label{eq:ppa} \\
q_\parallel &= \frac{\left(1+2\epsilon^2+\frac{1}{5}\epsilon^4\right)\left(1+\frac{\epsilon^2}{3}\right)}{(1+\epsilon^2)^2} \label{eq:qpa} \\
p_\bot &= 1 \nonumber \\
q_\bot &= \frac{\left(1+\frac{\epsilon^2}{3}\right)^2}{1+\epsilon^2} \nonumber
\end{align}
The reader might ask why we chose to use that specific form of the underlying distribution instead of the classic form of Ref.~\cite{MOT511} in which the individual components are gaussians, which might seem more satisfactory.
The point is that we need to account for all possible values of the pressure anisotropy, whereas a superposition of two isotropic gaussians is restricted to a positive anisotropy of the resulting distribution $\Ppa\ge\Ppe$).
Our choice meets this requirement, which is necessary to describe expanding plasmas, or the anisotropy arising in converging cylindrical geometry as is the case of hohlraums.
In addition, the calculations remain tractable, in contrast with other forms (such as the Pearson-IV distribution used, e.g., in Refs.~\cite{TOR10A,MIL162}).
However, we certainly agree that better forms of the closure distribution should be investigated, aiming at a better rendering of the fourth-order velocity moments.

\section{A plane one-dimensional implementation}
\label{sec:impl1D}

In the case of a plane one-dimensional (1D) situation where all quantities depend on the $x$-coordinate only, $\vec{\Omega}$ is the unit vector along the $x$ direction and the bulk velocity is $\vec{v}=v\vec{\Omega}$. Inserting Eqs. (\ref{eq:Pijaxis}), (\ref{eq:Qijkaxis}) and (\ref{eq:Rijkmaxis}) for this specific case, the 10-moment system (\ref{eq:drhodt})-(\ref{eq:dQdt}) thus reduces to the following six equations:
\begin{align}
\frac{d\rho}{dt} + \rho\frac{\partial v}{\partial x} &= 0 \label{eq:drhodt6} \\
\frac{dv}{dt} + \frac{1}{\rho}\frac{\partial \Ppa}{\partial x} &= \frac{F}{m} \label{eq:dvdt6} \\
\frac{d\Ppa}{dt} + 3\Ppa\frac{\partial v}{\partial x} + \frac{\partial \Qpa}{\partial x}&= \left(\frac{\partial \Ppa}{\partial t}\right)_c \label{eq:dPpadt6} \\
\frac{d\Ppe}{dt} + \Ppe\frac{\partial v}{\partial x}+\frac{\partial \Qpe}{\partial x} &= \left(\frac{\partial \Ppe}{\partial t}\right)_c \label{eq:dPpaedt6} \\
\frac{d\Qpa}{dt} + 4\Qpa\frac{\partial v}{\partial x} - 3\frac{\Ppa}{\rho}\frac{\partial \Ppa}{\partial x} + \frac{\partial \Rpapa}{\partial x} &= \left(\frac{\partial \Qpa}{\partial t}\right)_c \label{eq:dQpadt6} \\
\frac{d\Qpe}{dt} + 2\Qpe\frac{\partial v}{\partial x} - \frac{\Ppe}{\rho}\frac{\partial \Ppa}{\partial x} + \frac{\partial \Rpape}{\partial x} &= \left(\frac{\partial \Qpe}{\partial t}\right)_c \label{eq:dQpedt6}
\end{align}
where the fourth-order moment components $\Rpapa$ and $\Rpape$ are given by Eqs. (\ref{eq:Rpapa}) and (\ref{eq:Rpape}).

\subsection{Collisional relaxation terms}
\label{sec:relaxterms}

The relaxation terms on the r.h.s. of Eqs. (\ref{eq:dPpadt6})-(\ref{eq:dQpedt6}) are estimated from analytic values of the Coulomb collision frequency computed in known limit cases (see Appendix \ref{app:FP} and Appendix \ref{app:bimax}), and kinetic Fokker-Planck calculations as described in \cite{BAR107}.
From those items, a heuristic formula for the collisional relaxation time $\tau_c$ of the pressure anisotropy, such that
\[
\left(\frac{\partial \Ppa}{\partial t}\right)_c = \frac{P-\Ppa}{\tau_c},\quad \left(\frac{\partial \Ppe}{\partial t}\right)_c = \frac{P-\Ppe}{\tau_c}
\]
is designed, taking into account the actual features of the underlying velocity distribution.
The result is displayed on Fig. 5.
The important point here is that $\tau_c$ be given a realistic dependence on the anisotropy, reflecting the limit cases described in Sect. (\ref{sec:closplas}), rather than being kept constant.
More details are given in the following subsections.

For the present, proof-of-principle investigation, the same relaxation time has been used for the components of the heat flux tensor $\Qpa$ and $\Qpe$, but this might need to be improved.

\subsubsection{Analytic rates: the case of a bi-Maxwellian distribution}

In the case of a bi-Maxwellian distribution (with a Max\-wellian dependence in the longitudinal direction and in the transverse directions but with $P_\parallel\neq P_\bot$), an analytic value can be computed for the anisotropy relaxation rate (see Appendix \ref{app:bimax} and Refs. \cite{KOG611,SCH895}), assuming that the Maxwellian analytic form is preserved in the relaxation process (which is an approximation).
The result is
\begin{equation}
\frac{dP_\parallel}{dt} = \frac{P-P_\parallel}{\tau_{Max}(T)}F\left(\frac{P_\parallel-P}{P}\right) \label{eq:relKog}
\end{equation}
where $T=P/(nk_B)=(P_\parallel+2P_\bot)/(3nk_B)$ is the isotro\-pic part of the temperature ($n$ is the ion number density) and $\tau_{Max}$ is the collision time prevailing when the aniso\-tropy tends to vanish, defined by Eq. (\ref{eq:tauMbimax}).
$F(x)$ is a function which  tends to 1 for $x\to0$ (in the isotropic limit), but diverges for a large positive anisotropy ($x\to2$: ``cigar'' anisotropy). For a large negative anisotropy ($x\to -1$: ``pancake'' anisotropy), it takes on the finite limit value $\frac{5\pi}{2\sqrt{6}}$.
This function is displayed
{\setlength{\parindent}{0cm}
\parbox{\linewidth}{\small
%\rule{0pt}{2em}
\begin{center}
\includegraphics[scale=0.65]{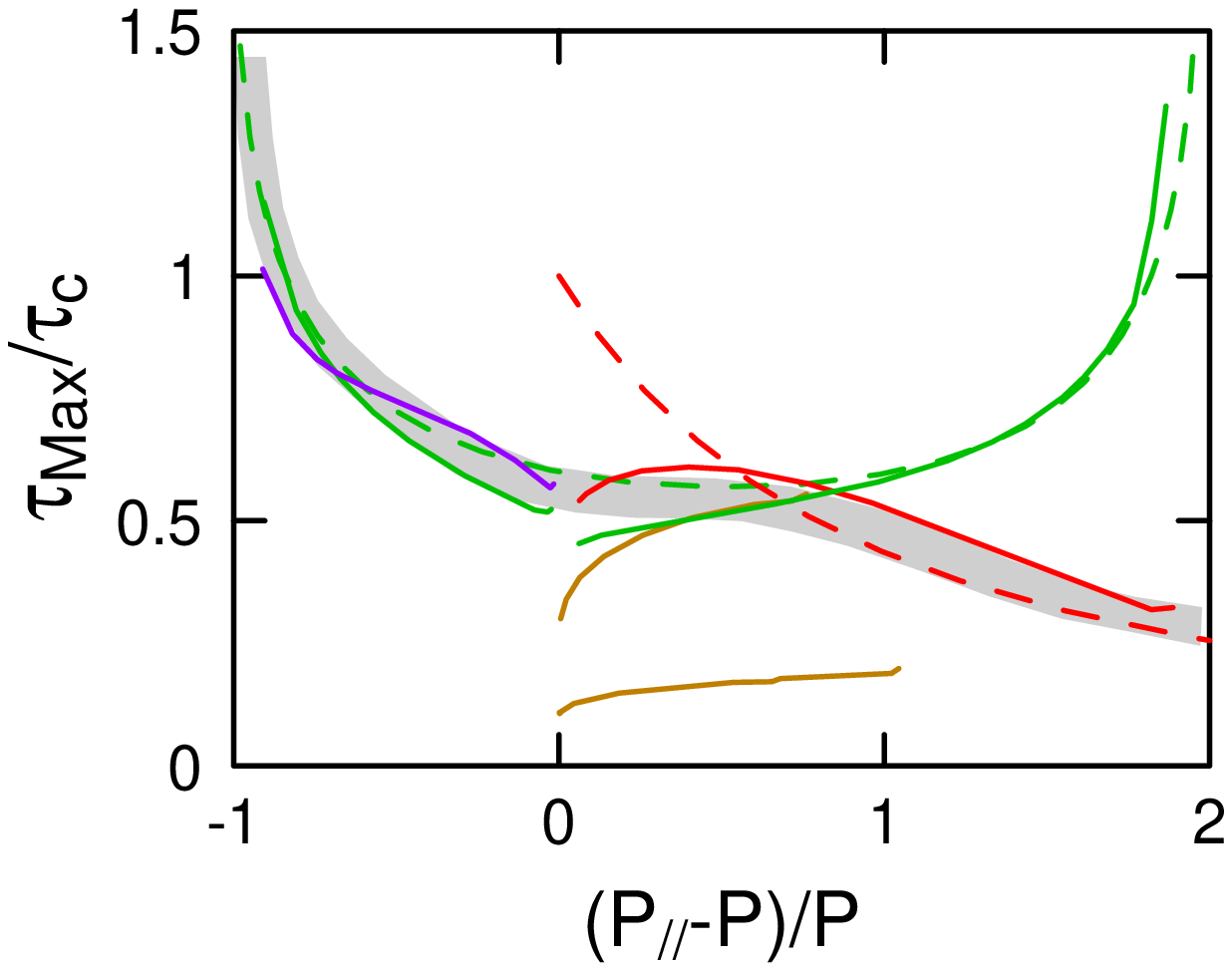}
\end{center}

\textbf{Fig. 5.}\rule{1em}{0pt}The ratio $\tau_{Max}/\tau_c$, where $\tau_c$ is the relaxation time of the pressure tensor anisotropy and $\tau_{Max}$ is the Maxwellian relaxation time for an interpenetrating-beam distribution close to isotropy, is plotted as a function of the relative pressure anisotropy $(\Ppa-P)/P$.
This is done for $\tau_c$ from known analytic formulas (dashed lines) and from numerical simulations of the Fokker-Planck equation (solid lines):
-- in red: for an initial distribution with two identical Maxwellian beams (the dashed curve displays the analytic value from Eq. (\ref{eq:Ffaisc})); % tracé avec plotf 0 2 -l -1 -b 0 -t 2 et la fonction eq:Ffaisc dans plotf
-- in brown: for initial distributions with two non-identical Maxwellian beams (for various sets of beam parameters, see Table 1);
-- in purple: for a toroidal initial distribution;
-- in green: for bi-Maxwellian initial distributions (the dashed curve displays the analytic value from \cite{KOG611}).
A heuristic value of the relaxation rate, valid for all values of the anisotropy in actual cases, is to be found in the grey region.

\rule{0pt}{2em}
}}
as the green dash\-ed curve on Fig. 5, however multiplied by a factor 0.6 because, for consistency, all results displayed on Fig. 5 are normalized to the same value of $\tau_{Max}$ corresponding to the interpenetrating-beam case discussed below.
In the following it is investigated whether the relaxation rate can be put into the simple form (\ref{eq:relKog}) in various physical situations.

\subsubsection{Analytic rates: the case of interpenetrating beams}

When the distribution is made of two interpenetrating beams in a 1D geometry, the relaxation rate can be estimated from the slowing-down model of Ref. \cite{LAR931}, which is itself validated by Fokker-Planck simulations.
The ion part of those equations is the following:
\begin{align}
\frac{dv_\alpha}{dt} &= -\nu_{\alpha\beta}(v_\alpha-v_\beta) \label{eq:relv} \\
\frac{dT_\alpha}{dt} &= \frac{2}{3}\nu_{\alpha\beta}\left(\frac{m}{2k_B}(v_\alpha-v_\beta)^2+T_\beta-T_\alpha\right) \label{eq:relT}
\end{align}
where $v_\alpha$ and $T_\alpha$ are the velocity and temperature of beam $\alpha$, $m$ is the ion mass and $\nu_{\alpha\beta}$ is a phenomenological collision frequency, for which Ref. \cite{BER917} gives an expression in the form:
\[
\nu_{\alpha\beta} = \frac{8\pi Z^4e^4\mbox{Log}\Lambda_{\alpha\beta}n_\beta} {m^2\left[(v_\alpha-v_\beta)^2+\zeta k_B(T_\beta+T_\alpha)/m\right]^{3/2}}
\]
$Z$ is the ionisation degree of the ions, $e$ is the elementary electric charge, $n_\beta$ is the number density (number of particles per unit volume) of beam $\beta$ and $\mathrm{Log}\Lambda_{\alpha\beta}$ is the Coulomb logarithm \cite{SIV661}.
Various values of the phenomenological coefficient $\zeta$ can be found in the literature, aiming at the best rendering of a kinetic simulation of the problem.
Ref. \cite{BER917} uses $\zeta=1$, Ref. \cite{DEC981} the value $\zeta=(9\pi/2)^{1/3}$ which is meant to reproduce exactly the two limit cases of a relative velocity small or large with respect to the thermal velocity.
A comparison is given on Fig. 21 of Ref. \cite{LAR931}.

\subsubsection{Modeling interpenetration by an anisotropy}

When the thermal energy per particle in the beams is small with respect to $m(v_\alpha-v_\beta)^2$ (i.e. in the limit of a large anisotropy for the global system), the above relaxation rate does not diverge as in the case of a bi-Maxwellian (see Fig. 5), but instead tends to a finite limit which depends on the relative beam velocity $v_\alpha-v_\beta$.
The behaviour described by Eqs. (\ref{eq:relv}-\ref{eq:relT}) can be translated in terms of a relaxation of $P_\parallel-P_\bot$ in the case of a ``cigar'' anisotropy ($P_\parallel>P_\bot$).
The pressure anisotropy for a two-beam distribution, assuming each beam to be isotropic (hence defined by a density $n_\alpha$, a velocity $v_\alpha$ and a pressure $P_\alpha=n_\alpha k_BT_\alpha$), reads:
\begin{align*}
\Ppe = (n_1+n_2)k_BT_\bot &= n_1k_BT_1+n_2k_BT_2 \\
\Ppa = (n_1+n_2)k_BT_\parallel &= n_1k_BT_1+n_2k_BT_2 \\
    & \hskip 3em + m\frac{n_1n_2}{n_1+n_2}(v_1-v_2)^2
\end{align*}
It is often legitimate to consider the beams individually isotropic, given the orders of magnitude of the various collision times (see Appendix \ref{app:FP}), as expressed for beam 1 in Eqs. (\ref{eq:therm}-\ref{eq:ralen}).
Taking this for granted, the above equations lead to
\[
\frac{d}{dt}(\Ppa-\Ppe) = 2m\frac{n_1n_2}{n_1+n_2}(v_1-v_2)\frac{d}{dt}(v_1-v_2)
\]
From Eq. (\ref{eq:relv}) we have
\[
\frac{d}{dt}(v_1-v_2) = -\frac{v_1-v_2}{\tau_R}
\]
with
\[
\tau_R = \frac{1}{\nu_{12}+\nu_{21}} = \frac {m^2\left[(v_1-v_2)^2+\zeta k_B(T_1+T_2)/m\right]^{3/2}}{8\pi Z^4e^4\mbox{Log}\Lambda_{12}(n_1+n_2)}
\]
so that
\[
\frac{d}{dt}(P_\parallel-P_\bot) = -2 \frac{P_\parallel-P_\bot}{\tau_R}
\]
$\tau_R$ is the relaxation time computed in \ref{app:FP}: Eq. (\ref{eq:tauRab}) inserting Eq. (\ref{eq:Tbstar}).
The same form as the Kogan formula \cite{KOG611} is found for the relaxation rate of the pressure anisotropy (see Eq. (\ref{eq:relKog})), with now:
\[
\tau_{Max} = \frac{3m^{1/2}(k_BT)^{3/2}}{8\sqrt{\pi}e^4Z^4\mbox{Log}\Lambda_{12}(n_1+n_2)}
\]
and in the case of equal-temperature beams ($T_1=T_2$)
\begin{equation}
F(x) = \left(1+\left(\left(\frac{3}{4\pi}\right)^{1/3}\frac{(n_1+n_2)^2}{n_1n_2}-1\right)\frac{x}{2}\right)^{-3/2}
\label{eq:Ffaisc}
\end{equation}
(the latter expression should be used only for $x\ge 0$).
This result is displayed, together with the Kogan formula for the bi-Maxwellian case, as the red dashed curve on Fig. 5.
A very different behaviour with res\-pect to the bi-Maxwellian formula is thus found for a large (positive) anisotropy, which is confirmed by the Fokker-Planck numerical simulation described below.

\subsubsection{Fokker-Planck simulations: the cases of a bi-Maxwellian and of two identical interpenetrating beams}

The following hypothesis is investigated here:
for whatever initial condition imposed on the ion distribution function, apart from very particular cases, the system will first undergo a transient stage where it quickly relaxes (at a rate depending on the specific initial condition chosen) to some ``universal'' distribution with the given pressure anisotropy, whose subsequent behaviour is then more or less the same in all cases.
This hypothesis is tested by means of Fokker-Planck numerical simulations of the collisional relaxation with the \textsc{fpion} code \cite{LAR931,LAR183}.
Various types of anisotropic initial conditions are used in this and the following subsections.

To begin with, an \textsc{fpion} case is initialised either with a bi-Maxwellian or with a two-beam distribution with the same global density and temperature (see parameters in Table 1), which is possible only for a positive anisotropy ($P_\parallel>P_\bot$). For $P_\parallel<P_\bot$, of course only bi-Maxwellian distributions can be set up.
For each of those cases, the relaxation rate of the relative pressure aniso\-tropy is monitored as the simulation proceeds.
The results are presented on Fig. 5, together with the analytic rates previously calculated as a comparison.
An animation of the distribution function $f(c_x,c_\bot,t)$ as time $t$ elapses during the relaxation process in the two-beam case is provided (see supplemental material: \texttt{animation1.gif}).

In the case of a bi-Maxwellian distribution, we essentially recover the results of Ref. \cite{SCH895}, namely that the numerically computed relaxation follows the analytic rate, although proceeding slightly more slowly as the system approaches isotropy, particularly for ``cigar'' anisotropy.
This can be attributed to the fact that large-velocity particles tend to relax more slowly due to the $\Delta v^3$ scaling of the collision time, which is expected to distort the distribution, whereas the analytic calculation assumes a bi-Maxwellian distribution for all values of the anisotropy.

In the case of the two-identical-beam distribution, the Fokker-Planck result agrees with the analytic rate for a large anisotropy, but then starts deviating from it for small to intermediate anisotropy. The relaxation rate measured in the simulation near isotropy is actually closer to the analytic value for the bi-Maxwellian case.
As the anisotropy relaxation proceeds along the solid red curve on Fig. 5, the distribution thus shifts from a metastable two-beam shape, as illustrated on Fig. 2c (for which the analytic rate displayed as the dashed red curve is valid), to a single component shape which is better described by a bi-Maxwellian (for which the analytic rate displayed as the dashed green curve is valid).
This behaviour is quite obvious on the provided animation of the distribution function $f(c_x,c_\bot,t)$ corresponding to the solid red curve (see supplemental material: \texttt{animation1.gif}).
The difference between the analytic rates for interpenetrating beams and the bi-Maxwellian near isotropy is due to the different analytic form for the distribution enforced in the case of interpenetration, even close to isotropy, which is thus not confirmed by the kinetic calculation.

\subsubsection{Fokker-Planck simulations: the relaxation of a toroidal anisotropy}

Formula (\ref{eq:Ffaisc}) is strictly valid only for a positive aniso\-tropy, because otherwise there is no way to split the distribution into well-separated components in velocity space while keeping azimuthal symmetry.
For example, in the case of a convergent
{\setlength{\parindent}{0cm}
\parbox{\linewidth}{
%\rule{0pt}{2em}
\parbox{\linewidth}{\small
\textbf{Table 1.} Parameters of the two-beam distributions used as initial conditions for \textsc{fpion} simulations of collisional relaxation: number density $n_\alpha$ in cm$^{-3}$, bulk velocity $v_\alpha$ in cm/s and temperature $T_\alpha$ in keV for beams $\alpha=1$ and 2, for two identical beams (id. b.) or non-identical beams (non-id. b. 1 and 2).}

\begin{center}
\begin{tabular}{l|ccc}
\hline
Case & id. b. & non-id. b. 1 & non-id. b. 2 \\
\hline
$n_1$ & $5\times10^{21}$ & $5\times10^{21}$ & $8\times10^{21}$ \\
$v_1$ & $1.17448\times10^8$ & $7.5\times10^7$ & $1.67754\times10^8$ \\
$T_1$ & 0.2 & 0.2 & 0.2 \\
$n_2$ & $5\times10^{21}$ & $5\times10^{21}$ & $2\times10^{21}$ \\
$v_2$ & $-1.17448\times10^8$ & $-7.5\times10^7$ & $-5\times10^7$ \\
$T_2$ & 0.2 & 5.88522 & 11 \\
\hline
\end{tabular}
\end{center}

\rule{0pt}{2em}
}}
 collision onto the axis in cylindrical geometry, on axis the velocity distribution is expected to assume the shape of a torus, and a specific expression must be calculated for the relaxation rate.
In the case of a central collision in spherical geometry, the velocity distribution at the centre is an isotropic shell, so that the pressure anisotropy vanishes.

A distribution modeling a toroidal anisotropy can be taken in the following form:
\begin{align*}
f(c_x,c_\bot) &= \frac{n}{B}\left(\frac{m}{2\pi k_BT_\parallel}\right)^{1/2}\frac{m}{2\pi k_BT_r} \times \\
              & \hskip 3em\mbox{exp}-\left(\frac{mc_x^2}{2k_BT_\parallel}+\frac{m(c_\bot-v_0)^2}{2k_BT_r}\right)
\end{align*}
with the normalization coefficient:
\[
B = e^{-u_0^2}+u_0\sqrt{\pi}(1+\mbox{erf}(u_0))
\]
where
\[
u_0 = \left(\frac{m}{2\pi k_BT_r}\right)^{1/2}v_0
\]
Taking the velocity moments of that distribution, we find that $n$ is the number density, $T_\parallel$ is the longitudinal temperature, and the perpendicular temperature is:
\[
k_BT_\bot = \frac{1}{2}\left(\left(3-\frac{e^{-u_0^2}}{B}\right)k_BT_r+mv_0^2\right)
\]
An \textsc{fpion} simulation is initialised with that distribution, taking $T_r=T_\parallel$ (the case of a torus with a circular section), and the result of the calculation is presented on Fig. 5.

\subsubsection{Fokker-Planck simulations: the case of two non-identical interpenetrating beams}

When the two beams have very different temperatures, the relaxation rate (as displayed on Fig. 5) remains different from the Maxwellian value when the anisotropy vanishes.
Taking a look at the velocity distribution function, this is due to the fact that the later part of the relaxation involves high-velocity particles from the hotter beam, with a larger collision time, while the distribution core, coming for the most part from the colder beam, has already relaxed.
The initialisation parameters for the cases presented on Fig. 5 are gathered in Table~1.

\subsubsection{Conclusion on the pressure anisotropy relaxation rate}\label{sec:impl1Dconc}

In summary, from the results gathered on Fig. 5 the following conclusions can be drawn:
\begin{itemize}
\item numerical results agree rather well with analytic expressions in the case of a bi-Maxwellian distribution, both for a positive (``cigar'') and a negative (``pancake'') anisotropy;
\item the relaxation of a toroidal distribution (supposedly typical of the situation on axis in the case of a collision in cylindrical geometry) follows the same path as a bi-Maxwellian distribution with the same anisotropy;
\item a distribution with two identical interpenetrating beams relaxes slightly faster than the analytic rate for intermediate values of the anisotropy;
this is due to the fact that the analytic rate is calculated assuming the beams to remain Maxwellian during the relaxation, which is not the case due to the different collision times for the core and non-thermal parts of the beam distributions;
for moderate to small values of the anisotropy at the end of the relaxation process, the distribution shifts to a one-component shape which relaxes slower than the two-beam analytic rate;
\item for interpenetrating beams with different temperatures (i.e., different widths in velocity space), the relaxation rate can become arbitrarily smaller than the Maxwellian value when isotropy is reached;
this is due to the fact that the high-velocity component from the hotter beam relaxes much more slowly (with a collision time scaling as $\approx \Delta v^3$), which leaves the distribution in a non-thermal state even after anisotropy has essentially vanished;
\item in all cases, the late stage of the numerical relaxation tends to proceed more slowly than expected from available analytic rates, which is attributed to distortions of the large-velocity part of the distribution which cannot be taken into account in analytic calculations.
\end{itemize}
From the latter facts, we will consider, in the parameter space $(\tau_{Max}/\tau_c,(P_\parallel-P)/P)$, a region (represented as the grey feature on Fig. 5) where the system is expected to stand in the most common physical situations.
We will thus leave aside, on one hand, the branch of bi-Maxwellian distributions with a strong positive anisotropy, which are not considered relevant because they need a particular ``preparation'' which is not expected to occur in the hohlraum plasmas we are studying, and on the other hand, interpenetrating-beam distributions with very different temperature components, which is a more serious problem.
This situation does occur, e.g., in delocalized electron thermal transport \cite{LUC853} in collisional plasmas, where it is known that the non-local features are essentially caused by the high-velocity, non-thermal part of the electron distribution function.
But this is a different physical problem, relevant for high-gradient quasi-stationary situations which prevail on longer interaction times.

We will also not be able to describe the slower rates found in the later part of the relaxation, attributed to distortions of the high-velocity part of the distribution which would need a more complete moment description including a dynamical treatment of kurtosis (related to moments of order 4).

In the present, short interaction time, plasma collision situations, as far as ion-distribution features are concerned we will thus use a heuristic fit of the relaxation time which renders the grey region on Fig. 5.

\subsection{Hyperbolicity}
\label{sec:hyperb}

Before implementing Eqs. (\ref{eq:drhodt6}-\ref{eq:dQpedt6}) in a hydrodynamics code, we have to check whether the expected solutions are stable, which, from the mathematical point of view, is related to the hyperbolicity of the system.
Depending on the specific closure used, some linear modes of the system might be amplified, in relation with the occurrence of imaginary solutions of their dispersion equation, for some sets of values of the parameters used for the closure.
This is indeed the case for the classic 13-moment system of Grad \cite{GRA491}, which is found to be stable only in a limited region of the parameter space defined by the pressure anisotropy and the heat flux \cite{TOR001}.

We thus linearize Eqs. (\ref{eq:drhodt6}-\ref{eq:dQpedt6}) for perturbations of high frequency $\omega$ and large wave vector $\vec{k}$.
All quantities are taken in the form $Q+Q_1\exp(i(kx-\omega t))$ where $Q_1$ is the complex amplitude of the supposedly small perturbation.
From the high-frequency assumption, variations of the unperturbed quantities $\partial Q/\partial t$ and $\partial Q/\partial x$ are neglected with respect to variations of the perturbation terms $i\omega Q_1$ and $ikQ_1$.
The resulting dominant terms, defining $\lambda=\lb$, are presented in Eq. (\ref{eq:syshyp}).
\begin{table*}
\begin{equation}
\left(\begin{array}{cccccc}
\poto\lambda                       & -1                       & 0                                        & 0                             & 0                    & 0  \\
\poto 0                            & \lambda                  & -1                                       & 0                             & 0                    & 0  \\
\poto 0                            & \frac{-3\Ppa}{\rho}      & \lambda                                  & -1                            & 0                    & 0  \\
\frac{p_\parallel\Ppa^2}{\rho^2}   & \frac{-4\Qpa}{\rho}      & \left(\frac{(3-2p_\parallel)\Ppa}{\rho}
                                                   +\frac{q_\parallel\Qpa^2}{\Ppa^2}\right) & {\left(\lambda-\frac{2q_\parallel\Qpa}{\Ppa}\right)}
                                                                                                                            & 0                    & 0  \\
\poto 0                            & {\gm\frac{-\Ppe}{\rho}}  & 0                                        & 0                             & \lambda              & -1 \\
{\gm\frac{p_\bot\Ppa\Ppe}{\rho^2}} & {\gm\frac{-2\Qpe}{\rho}} & {\gm\left(\frac{q_\bot\Qpa\Qpe}{\Ppa^2}
                                                       +\frac{(1-q_\bot)\Ppe}{\rho}\right)} & {\gm\frac{-q_\bot\Qpe}{\Ppa}} & {\frac{-p_\bot\Ppa}{\rho}}
                                                                                                               & {\left(\lambda-\frac{q_\bot\Qpa}{\Ppa}\right)}
\end{array}\right)
\left(\begin{array}{c}
\poto\rho_1 \\
\poto \rho v_1 \\
\poto\Pupa \\
\poto\Qupa \\
\poto P_{1\bot} \\
\poto Q_{1\bot}
\end{array}\right)=\left(\begin{array}{c}\poto 0 \\ \poto 0 \\ \poto 0 \\ \poto 0 \\ \poto 0 \\ \poto 0 \end{array}\right) \label{eq:syshyp}
\end{equation}
\end{table*}
The r.h.s. of Eq. (\ref{eq:syshyp}) vanishes because the collision times are assumed much larger than the perturbation period $1/\omega$.
Anyway the collision terms, if not negligible, are expected to damp the wave perturbations, thus stabilizing the system, rather than enhancing non-hyperbolicity.

The determinant of that system is the product of a ``transverse'' factor
\[
\lambda\left(\lambda-\frac{q_\bot\Qpa}{\Ppa}\right) - \frac{p_\bot\Ppa}{\rho}
\]
with the two roots
\[
\lambda = \frac{q_\bot\Qpa}{2\Ppa} \pm \left(\left(\frac{q_\bot\Qpa}{2\Ppa}\right)^2+\frac{p_\bot\Ppa}{\rho}\right)^{1/2}
\]
which are real for all values of the parameters $\Ppa$ and $\Qpa$, and a ``longitudinal'' factor which reads, defining $x=\lambda(\rho/\Ppa)^{1/2}$ and $\xi=\rho^{1/2}\Qpa/\Ppa^{3/2}$:
\[
D(x) = x^4-2q_\parallel\xi x^3+(q_\parallel\xi^2-2p_\parallel)x^2+2(3q_\parallel-2)\xi x+p_\parallel
\]
The nature of the roots of the dispersion equation \linebreak$D(x)=0$ depends on the specific closure chosen (by means of the numerical parameters $p_\parallel$ and $q_\parallel$) and the physical parameter $\xi$ which is the ratio of the parallel heat flux to its free-streaming value.
They will be studied in the case $\xi\ge 0$ because any root for $\xi < 0$ is the opposite of a root for $\xi\ge 0$, so that hyperbolicity (i.e., the question whether all roots are real or occur in complex-conjugate pairs) depends only on $|\xi|$.
The following properties are easily proven:
\begin{itemize}
\item for all closures, hyperbolicity prevails in a non-empty vicinity of $\xi=0$; this is due to the fact that the dispersion equation, which in this case reads $x^4-2p_\parallel x^2+p_\parallel=0$, has four simple real roots, at least for $p_\parallel>1$, which is the generic case considered here (see Eq.~(\ref{eq:ppa}));
the roots depend continuously on $\xi$, so that this property should remain true for small, but non-vanishing values of $\xi$;
\item in the case of the double-foil closure ($p_\parallel=q_\parallel=1$, see Eqs. (\ref{eq:ppa}) and (\ref{eq:qpa})), the dispersion equation reads $D(x)=(x^2-\xi x -1)^2=0$, with four real roots for all $\xi$.
\item For $|\xi|\to\infty$, the left minimum of the dispersion function is at $x=x_{min} \sim \frac{2-3q_\parallel}{q_\parallel\xi}$, and the corresponding value is $D(x_{min}) \sim p_\parallel-\frac{(2-3q_\parallel)^2}{q_\parallel}$.\\
The hyperbolicity condition is thus $p_\parallel\le\frac{(2-3q_\parallel)^2}{q_\parallel}$, which is not met for the general double-waterbag closure.
However, a numerical exploration % pour epsilon = 1, semble hyperbolique pour xi <= 1.4732397
shows that the corresponding growth rate (imaginary part of the root) remains small, so that it might be easily compensated for by other effects in the complete physical model (e.g., damping by electron-ion collisions).
In addition, it was found that in very low-density regions the ratio of the heat-flux components to the free-streaming flux, or in other words the above-defined parameter $\xi$, had to be limited.
Hence this loss of hyperbolicity is maybe not an issue in actual simulations.
\end{itemize}

\section{Numerical scheme and results}
\label{sec:numres}

\subsection{A plane one-dimensional numerical implementation}

The 6-moment system with ``double-waterbag'' closure described above has been implemented in the \textsc{multif} code \cite{CHE979}, in the limit of a plane 1D geometry.
The numerical scheme used is derived from the ``central-upwind'' formalism of Kurganov and Lin \cite{KUR072,KUR161}, with special care for the compatibility between the transported moments.
The main features of the scheme are as follows.

We use a finite-volume formalism for a system of hyperbolic conservation laws, with collisional relaxation and the interaction with the electron fluid (including acceleration by the electric field and electron thermal conduction) treated by a split-step strategy.
As in Refs. \cite{KUR072,KUR161}, each discretization mesh is decomposed into ``staggered'' and ``regular'' sub-cells, over which the usual numerical procedure is constructed, namely:
\begin{itemize}
\item a reconstruction step of the moment-vector profile using a slope-limiting algorithm;
\item an evolution step taking into account the convective terms of the system; in the present case a kinetic implementation \cite{PER92A,EST962} of the convective fluxes is used, ensuring that the fluxes are consistent with the underlying closure distribution function;
\item a projection step of the modified moments back onto the numerical grid, using a slope limiting algorithm.
\end{itemize}
The three steps just described can be reformulated as a diffusion---convection---antidiffusion scheme, which displays an interesting analogy with the Flux-Corrected Transport (FCT) paradigm \cite{BOR732,KUZ123}.
The slope-limiting algorithm used includes a further limiting step to ensure the admissibility and compatibility of the interpolated moments, in the spirit of \cite{VAN982,LIS114}.
The final expression of the scheme is a set of fluxes between the spatial grid cells, which explicitly conserve the moments, and provide a consistent basis for the advection of possible additional degrees of freedom or passive quantities such as the ionisation degree.

Due to their small mass (or to their high plasma frequency $\omega_p=\left(\frac{4\pi n_ee^2}{m_e}\right)^{1/2}$), electrons are treated as a neutralizing fluid with density $n_e$, velocity $v_e$ and temperature $T_e$, taking into account the collisional interaction with the ions through the usual friction and temperature relaxation coefficients, as described in Appendix %\linebreak 
\ref{app:FP}.
From the quasi-neutrality assumption we have $n_e=Zn$ and in a 1D geometry $v_e=v$.
The temperature is governed by the second velocity moment of the electron kinetic equation, which reads
\begin{align*}
& \frac{\partial}{\partial t}\left(\frac{3}{2}n_ek_BT_e\right) + \frac{\partial}{\partial x}\left(\frac{3}{2}v_en_ek_BT_e\right) \\
&  \hskip 3em + n_ek_BT_e\frac{\partial v_e}{\partial x} + \frac{\partial q_e}{\partial x}
   = \frac{3}{2}n_e\left(\frac{\partial k_BT_e}{\partial t}\right)_c
\end{align*}
where $q_e$ is the electron heat flux and the r.h.s. accounts for the collisional interaction with the ions.
In addition, the assumption of a small electron mass leads to the following value of the electric field:
\[
eE_x = -\frac{1}{n_e}\frac{\partial(n_ek_BT_e)}{\partial x}
\]

More details about the numerical scheme will be given in a separate paper.

\subsection{Numerical results: the collision of two plasmas in plane geometry}

To check the ability of the extended model to account for hydrodynamic quantities in cases where plasma interpenetration is expected to occur, the situation described in Table 2 and Fig. 6 was used as an initial condition.
Namely, at time $t=0$ two plasma slabs are initially drifting towards each other at a relative velocity $|v_2-v_1|=2.25\times 10^8$ cm/s, much larger than their thermal velocity $\left(\frac{k_BT_i}{Am_p}\right)^{1/2}\approx 2.2\times 10^6$ cm/s.
Those parameters are such that the slowing-down distance through
{\setlength{\parindent}{0cm}
\parbox{\linewidth}{\small
\rule{0pt}{2em}
\begin{center}
\includegraphics[scale=0.45]{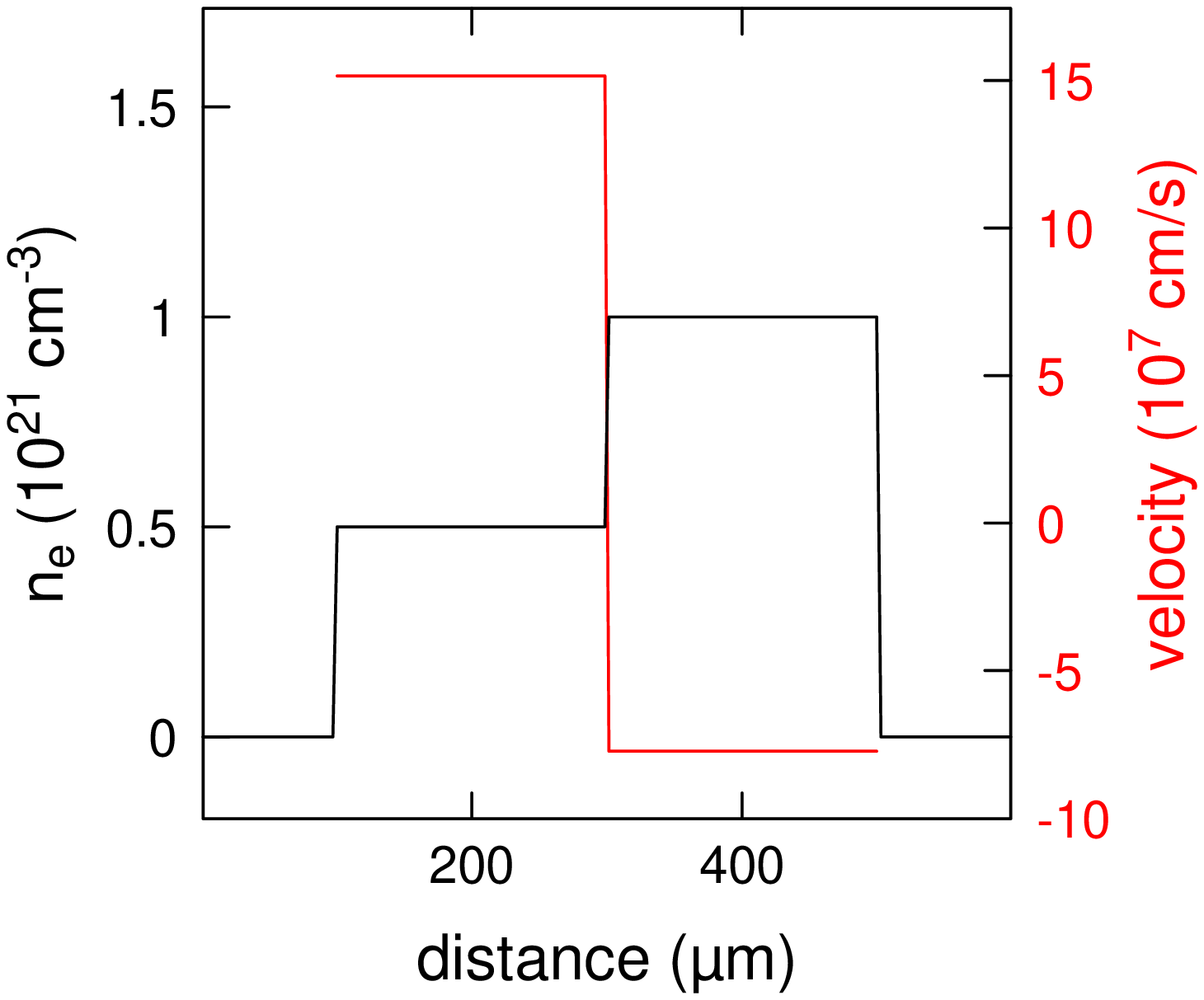}
\end{center}

\textbf{Fig. 6.}\rule{1em}{0pt}Profiles of the electron density $n_e=Z_in_i$ and bulk velocity $v_i$ at time $t=0$ for the plasma collision test case.
}}
ion-ion collisions of the two plasmas $\lambda_{12} = |v_2-v_1|\tau_{R12}\approx 1$ mm, where the slowing-down time $\tau_{R12}$ is given by Eq. (\ref{eq:tauRab}), is long enough that the slabs should pass through each other.
The results shown in the following were obtained using the ``double-waterbag'' model distribution described in Sect. \ref{sec:axisclos} with $\epsilon=1$ (the two components of the distribution are contiguous in velocity space).
The two plasma slabs have different densities, so that the asymmetry in their interpenetration will translate into a strong heat flux, thus providing a test of the present hydrodynamics model, extended to velocity moments of order 3.
Four types of calculation were performed from the above initial condition, with various modeling options, the results of which are gathered on Fig. 7, namely:
\begin{itemize}
\item standard single-fluid (Fig. 7a): an Euler-equation behaviour is simulated by cancelling the higher-order moments (pressure anisotropy $\Rightarrow \Ppa=\Ppe$ and heat flux components $\Rightarrow \Qpa=\Qpe=0$) after every time step, which models an instantaneous relaxation through very strong collision rates $\tau_c\to 0$;
\item multifluid (Fig. 7d): the two plasma slabs are treated as separate fluids, and allowed to drift through each other while undergoing the usual effects of friction and temperature relaxation; this is considered the reference solution of the problem;
\item single-fluid, extended to order 2 (Fig. 7b): the hydrodynamics model is extended beyond Euler equations by including a dynamical treatment of the pressure tensor components $\Ppa\neq\Ppe$, with a vanishing heat-flux closure ($\Qpa=\Qpe=0$); this is the legacy model implemented in \textsc{multif} \cite{CHE979}, but here realistic collisional relaxation rates are used, as discussed above;
\item single-fluid, extended to order 3 (Fig. 7c): this is the present new model, including a dynamical treatment of the pressure tensor components $\Ppa\neq\Ppe$ and heat-flux tensor components $\Qpa\neq 0$, $\Qpe\neq 0$, with the interpenetration-like closure described above.
\end{itemize}
The results are displayed on Fig. 7 at time $t=75$ ps, when the plasma slabs have come to almost complete interpenetration in the two-fluid model.
As expected, the reference
{\setlength{\parindent}{0cm}
\parbox{\linewidth}{
\rule{0pt}{1.7em}

\parbox{\linewidth}{\small
\textbf{Table 2}. Initial condition for the plasma collision test case. The charge state $Z_i$, the atomic mass number $A_i$, the ion number density $n_i$ (cm$^{-3}$), the bulk velocity $v_i$ (cm/s), and the ion and electron temperatures $T_i$ and $T_e$ (keV) are given for each plasma slab $i=1,2$.}

\begin{center}
\begin{tabular}{ccccccc}
\hline
$i$ & $Z_i$ & $A_i$ & $n_i$            &     $v_i$          & $T_i$ & $T_e$ \\
\hline
1   &   50  &  197  &     $10^{19}$    & $1.5\times10^{8}$  &   1   &   1   \\
2   &   50  &  197  & $2\times10^{19}$ & $-7.5\times10^{7}$ &   1   &   1   \\
\hline
\end{tabular}
\end{center}
%\rule{0pt}{1em}
}}
two-fluid simulation displays (see Fig. 7d) a square-shaped density profile resulting from the superposition of the interpenetrated plasmas.
The profile edges are slightly smoothed due to self-expansion of the plasmas into the surrounding vacuum.
The components of the pressure tensor exhibit almost flat profiles, with a very strong anisotropy ($\Ppa\gg\Ppe$) due to the contribution from the square of the large relative velocity between the two streams.

\begin{figure*}
\begin{center}
\includegraphics[scale=0.38]{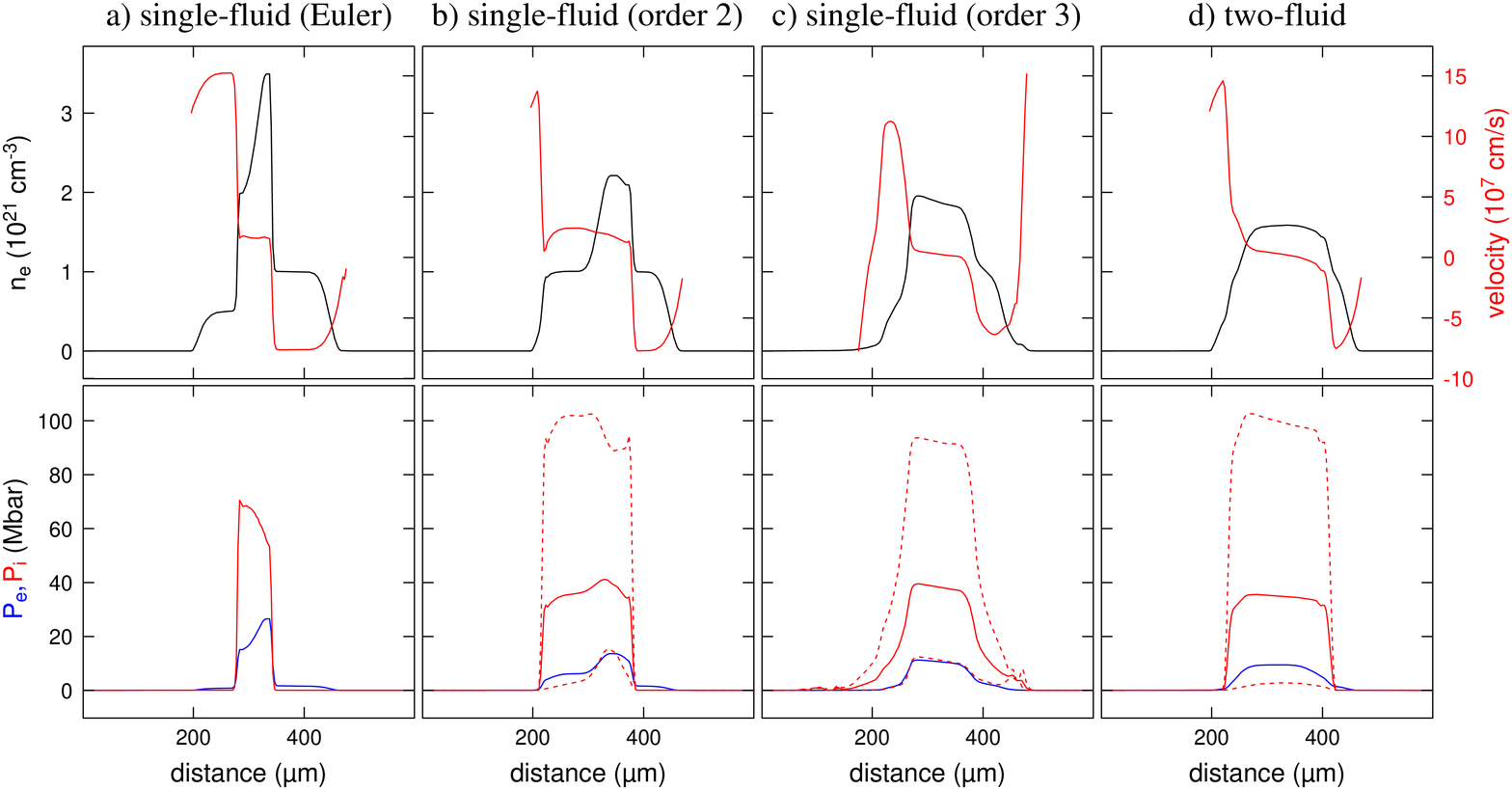}
\end{center}

{\small\textbf{Fig. 7.}\rule{1em}{0pt}Profiles of hydrodynamic quantities at time $t=75$ ps, obtained with four different modeling options (see text). Top row: electron density $n_e=Z_in_i$ (in black) and global bulk velocity (in red). Bottom row: electron pressure $P_e$ (in blue) and components of the global ion pressure tensor (red dotted lines) $\Ppa$ and $\Ppe$ ($\Ppa$ is the largest); the scalar pressure $P_i=(\Ppa+2\Ppe)/3$ is also displayed (red solid lines).
}
\end{figure*}
In marked contrast, the standard single-fluid Euler-like calculation (see Fig. 7a) displays the expected features in that

Riemann problem, namely two strong shock waves propagating away from the central contact discontinuity arising from the initial density jump between the two slabs.
The pressure profiles display a vanishing anisotropy, and have propagated much more slowly than the edges of the corresponding profiles in the two-fluid interpenetration case.
As a result of the shock compression, a strong density ridge has built up in the denser slab, which is of course not present in the two-fluid simulation.
This non-physical feature is hopefully expected to disappear with a more appropriate treatment.

In the single-fluid calculation taking into account pressure anisotropy, but not the third-order moments, (see Fig. 7b), the pressure profiles have come to a much better agreement with their two-fluid counterpart, as regards both the amplitude of the anisotropy and the location of the shock fronts.
However, the density overshoot at the contact discontinuity, although smaller, is still there.
Let us mention that in the case of the collision of equal-density slabs, we checked that the contact discontinuity disappears as expected, so that all hydrodynamic profiles agree with the two-fluid simulation.
This is due to the fact that in this case the global velocity distribution of the interpenetrating plasmas is symmetric, so that moments of order 3 vanish, in agreement with the closure hypothesis.

Finally, in the single-fluid calculation with our new model, taking also into account the components of the heat flux (see Fig. 7c), the contact discontinuity has disappeared as expected, although the pressure profiles are not as satisfactory as those of Figs. 7b and 7d.
This is an encouraging result as a proof-of-principle demonstration that higher-order moment models can hopefully treat interpenetration situations in a satisfactory way, although some improvements are obviously still needed, as discussed in the final section below.

\section{Conclusion and paths forward}
\label{sec:conclu}

A new extended hydrodynamics model resting on a set of 10 moments of the velocity distribution function is presented.
The model targets plasma collision and/or interpenetration situations which challenge the numerical simulation of ICF hohraums.
For this purpose, a closure is designed with the ability of describing the double-humped velocity distributions which can arise in a plasma submitted to a strong collision or interpenetration.
This leads to a specific set of moments, different from the usual 10-moment or 13-moment descriptions, namely: density, bulk velocity, Euler angles of the symmetry axis, parallel and perpendicular pressure, parallel and perpendicular heat flux.
The properties of the model (realizability, hyperbolicity) are studied in detail.

The new model was implemented in a reduced, plane 1D geometry, using a specific numerical scheme which will be presented elsewhere.
An academic test case is presented involving the high-velocity collision of two plasma slabs (comprising a single ion species), and the solution is compared with the results of other models, including single-fluid Euler-equa\-tion hydrodynamics, and interpenetrating multi-fluid hydrodynamics.
The new model is found to satisfactorily reproduce the main features of the multifluid simulation, without the non-physical density ridges produced by the classical single-fluid model.
This demonstrates that a higher-order moment model is a valid option for the simulation of non-equilibrium features of the ion species in colliding plasmas.
The other step needed for a complete simulation of plasma interpenetration is to integrate this model for the treatment of each species in a multi-species, multifluid code \cite{MAR212}.

To go beyond this proof-of-principle result, some improvements and further developments are in order.
These include
\begin{itemize}
\item a more detailed study of the realizability and hyperbolicity of the model, particularly in two- or three-dimensional (2D or 3D) geometry;
\item an improved closure, hopefully reproducing the Max\-wellian limit at isotropy (e.g., in the spirit of the Pearson-IV closure \cite{TOR10A,MIL162}), or at least with a better rendering of the kurtosis;
\item a specific relaxation rate for moments of order 3 (different from those for moments of order 2);
\item as shown on Fig. 5 and as discussed in Sect. \ref{sec:impl1Dconc}, it might be useful to treat the relaxation of the non-thermal, high-velocity part of the distribution by a still higher-order moment model, including a dynamic treatment of kurtosis (moments of order 4);
\item the application of the model to non-ideal plasmas and its compatibility with the use of an equation of state for the scalar pressure should be investigated (however keeping in mind that the kinetic effects investigated here only occur in tenuous hot plasmas, and are expected to merely disappear in denser media with smaller collision times).
\end{itemize}
More specifically, as regards the 1D implementation of the model in multifluid codes such as \textsc{multif}, the following points are in order:
\begin{itemize}
\item a finalization of the numerical scheme, and its description in a dedicated paper;
\item the implementation of spherical geometry, which\linebreak would allow the model to be used for the simulation of exploding-pusher ICF implosions \cite{LAR183};
\end{itemize}
For a 2D or 3D implementation, a key point to address is the handling of the Euler angles defining the symmetry axis $\vec{\Omega}$ of the closure distribution.
An axisymmetric velocity distribution is of course not a generally valid description of a 2D or 3D situation, but our model rests on the assumption that it is generically valid, meaning that it will account for most of the volume of the system under investigation, where the collision or interpenetration of flows is locally 1D.
The question of implementing such an N-moment model in large radiative-hydrodynamics codes \cite{HIG216,ZIM752} together with a multi-species, multifluid capability will then be open.

\appendix

\section{Fokker-Planck relaxation times for Coulomb collisions}
\label{app:FP}
The expressions of the collision times used in the main part of the paper are derived here from the Fokker-Planck equation governing Coulomb collision processes, in the reference form of Rosenbluth {\itshape et al.} \cite{ROS573}.
The term governing the evolution of the velocity distribution function $f_a(\vec{c})$ for particles of species $a$ due to collisions on particles of species $b$ is:
\begin{equation}
\left(\frac{\partial f_a}{\partial t}\right)_{a\rightarrow b} = -\frac{\partial J_{(ab)i}}{\partial c_i} \label{eq:dfdtcab}
\end{equation}
where the current in velocity space $\mathbf{J}_{(ab)}$ is the sum of a convection term and a diffusion term:
\begin{equation}
J_{(ab)i} = 4\pi\Gamma_{ab}\left(-\frac{m_a}{m_b}\frac{\partial \mathcal{S}_b}{\partial c_i}f_a + \frac{\partial^2\mathcal{T}_b}{\partial c_i\partial c_j}\frac{\partial f_a}{\partial c_j}\right) \label{eq:JFP}
\end{equation}
with the Rosenbluth potentials $\mathcal{S}_b$ and $\mathcal{T}_b$ defined by:
\begin{equation}
\Delta_c\mathcal{S}_b = f_b \quad ; \quad \Delta_c\mathcal{T}_b = \mathcal{S}_b \label{eq:Rpots}
\end{equation}
and with
\[
\Gamma_{ab} = \frac{4\pi e^4Z_a^2Z_b^2}{m_a^2}\mathrm{Log}\Lambda_{ab}
\]
where $Z_se$ and $m_s$ are the charge and mass of particles of species $s=a,b$ and $\mathrm{Log}\Lambda_{ab}$ is the Coulomb logarithm defined, e.g., in Ref. \cite{SIV661}.
We use a definition of the potentials slightly different from the original $h_b$ and $g_b$ from reference \cite{ROS573}; the correspondence is:
\[
\mathcal{S}_b = -\frac{m_b}{m_a+m_b}\frac{h_b}{4\pi} \quad ; \quad \mathcal{T}_b = -\frac{g_b}{8\pi}
\]
A general integral expression of the Rosenbluth potentials, which is a solution of the Poisson equations (\ref{eq:Rpots}), is
\begin{align}
S_b(\vec{c}) &= \frac{-1}{4\pi}\int \frac{f_b(\vec{c}^\prime)}{|\vec{c}-\vec{c}^\prime|}\mbox{d}^3c^\prime \label{eq:Sb}\\
T_b(\vec{c}) &= \frac{-1}{8\pi}\int f_b(\vec{c}^\prime)|\vec{c}-\vec{c}^\prime|\mbox{d}^3c^\prime \label{eq:Tb}
\end{align}
More useful expressions can be obtained in specific cases.

\paragraph{The case of a Maxwellian distribution for target particles.}

If $f_b$ is the Maxwellian:
\[
f_b(\vec{c}) = n_b\left(\frac{m_b}{2\pi k_BT_b}\right)^{3/2}e^{-u^2}
\]
with
\[
\vec{u} = \left(\frac{m_b}{2k_BT_b}\right)^{1/2}(\vec{c}-\vec{v}_b)
\]
where $n_b$, $\vec{v}_b$ and $T_b$ are the density, bulk velocity and temperature of species $b$, then the Rosenbluth potentials can be explicitly computed:
\begin{align*}
\mathcal{S}_b(\vec{c}) &= -\frac{n_b}{4\pi}\left(\frac{m_b}{2k_BT_b}\right)^{1/2}\frac{\mbox{erf}(u)}{u} \\
\mathcal{T}_b(\vec{c}) &= -\frac{n_b}{8\pi}\left(\frac{2k_BT_b}{m_b}\right)^{1/2}\left(\frac{e^{-u^2}}{\sqrt{\pi}} + \left(u+\frac{1}{2u}\right)\mbox{erf}(u)\right)
\end{align*}
For electrons, to lowest order in powers of the electron/ion mass ratio, we get:
\begin{align*}
\frac{\partial \mathcal{S}_e}{\partial c_i} & = \frac{n_e}{3\pi^{3/2}}\left(\frac{m_e}{2\tau_e}\right)^{3/2}(c_i-v_{e,i}) \\
\frac{\partial^2\mathcal{T}_e}{\partial c_i\partial c_j} & = -\frac{n_e}{6\pi^{3/2}}\left(\frac{m_e}{2\tau_e}\right)^{1/2}\delta_{ij}
\end{align*}
so that the electron collision term for ions of species $a$ finally reads:
\begin{align*}
\left(\frac{\partial f_a}{\partial t}\right)_{ae} &= \frac{4\sqrt{2\pi}e^4Z_a^2m_e^{1/2}n_e}{3m_a(k_BT_e)^{3/2}}\mathrm{Log}\Lambda_{ae} \\
 & \hskip 5em \times\frac{\partial}{\partial c_i}\left((c_i-v_{e,i})f_a + \frac{k_BT_e}{m_a}\frac{\partial f_a}{\partial c_i}\right)
\end{align*}
For collisions on ions, we get:
\begin{align}
\frac{\partial \mathcal{S}_b}{\partial c_i} & = \frac{n_b}{3\pi^{3/2}}\left(\frac{m_b}{2k_BT_b}\right)^{3/2}R(u)(c_i-v_{b,i}) \label{eq:dSdv} \\
\frac{\partial^2\mathcal{T}_b}{\partial c_i\partial c_j} & = -\frac{n_b}{6\pi^{3/2}}\left(\frac{m_b}{2k_BT_b}\right)^{1/2} \nonumber \\
   & \hskip 4em \times\left[\left(\delta_{ij}-\frac{u_iu_j}{u^2}\right)L(u) + \frac{u_iu_j}{u^2}R(u)\right] \nonumber
\end{align}
where the following functions have been defined:
\begin{align*}
R(u) & = \frac{3}{2u^2}\left(\frac{\sqrt{\pi}\mathrm{erf}(u)}{2u}-e^{-u^2}\right) \\
 & \buildlow{\sim}{0} 1-\frac{3u^2}{5} \dots \\
 & \buildlow{\sim}{\infty} \frac{3\sqrt{\pi}}{4u^3} \\
L(u) & = \frac{3}{4u^2}\left(e^{-u^2}+\left(2u-\frac{1}{u}\right)\frac{\sqrt{\pi}}{2}\mathrm{erf}(u)\right) \\
 & \buildlow{\sim}{0} 1-\frac{u^2}{5} \dots \\
 & \buildlow{\sim}{\infty} \frac{3\sqrt{\pi}}{4u}
\end{align*}
To get synthetic formulas for the orders of magnitude of collision times, we will use a more practical approximation of $R(u)$, which reproduces those two limits:
\[
R(u) \rightarrow \left(1+\left(\frac{2}{9\pi}\right)^{1/3}2u^2\right)^{-3/2}
\]

\paragraph{Definition of a slowing-down time.}

From the expression (\ref{eq:JFP}) of the current in velocity space, the slowing-down rate of distribution $a$ by target particles $b$ is obtained:
\begin{align}
\frac{dv_{a,i}}{dt} & = \frac{1}{n_a}\int c_i \left(\frac{\partial f_a}{\partial t}\right)_{a\rightarrow b}d^3c \nonumber \\
 & = -\frac{4\pi\Gamma_{ab}}{n_a}\left(1+\frac{m_a}{m_b}\right)\int\frac{\partial \mathcal{S}_b}{\partial c_i}f_a(\mathbf{c})d^3c \label{eq:DVDT}
\end{align}
Among the two terms in the latter expression, the second one comes directly from the ``slowing-down'' term in the Fokker-Planck equation, and the first one comes from the variation of the diffusion tensor inside the region where $f_a$ takes on non-negligible values.
Let us notice that so far, no approximation was made, and that expression is exact;
in particular it conserves momentum in $a\leftrightarrow b$ collisions, since it is in the form $1/n_am_a$ $\times$ a factor which is symmetric in the exchange $a\leftrightarrow b$ $\times$ the integral which is antisymmetric in the exchange $a\leftrightarrow b$, as can be seen after three integrations by parts, taking into account that $\Delta_cS_a = f_a$.
We thus obtain
\[
n_am_a\frac{dv_{a,i}}{dt} + n_bm_b\frac{dv_{b,i}}{dt} = 0
\]
When the distribution $f_a$ is very localized (very cold), in the integral (\ref{eq:DVDT}) we can factor out $\partial\mathcal{S}_b/\partial c_i$.
But then, the symmetry which leads to the explicit momentum conservation is broken, because a further assumption was made about $f_a$ with respect to $f_b$.
If we further assume that $f_b$ is Maxwellian, inserting Eq. (\ref{eq:dSdv}), we obtain:
\begin{align}
\frac{dv_{a,i}}{dt} &= -4\pi\Gamma_{ab}\left(1+\frac{m_a}{m_b}\right) \frac{n_b}{3}\left(\frac{m_b}{2\pi k_BT_b}\right)^{3/2} \nonumber \\
   & \hskip 9em \times R(u)(v_{a,i}-v_{b,i}) \label{eq:DVDTloc}
\end{align}
where we recall that:
\[
u = \left(\frac{m_b}{2k_BT_b}\right)^{1/2}|\vec{v}_a-\vec{v}_b|
\]
The exact expression of $R(u)$ was derived above in the case of a Maxwellian $f_b$.
If in addition $f_b$ is also localized (with a thermal velocity $k_BT_b/m_b$ much smaller than the relative velocity $\vec{v}_a-\vec{v}_b$), then the lost symmetry is recovered, since we know (see above) that in that case
\[
R(u) \rightarrow \frac{3\sqrt{\pi}}{4u^3}
\]
so that the slowing-down rate reads:
\[
\frac{dv_{a,i}}{dt} = -n_b\Gamma_{ab}\left(1+\frac{m_a}{m_b}\right) \frac{v_{a,i}-v_{b,i}}{|\vec{v}_a-\vec{v}_b|^3}
\]
We can check that the above expression is in the form $1/n_am_a$ $\times$ a term which is antisymmetric in the exchange $a\leftrightarrow b$.
To recover the symmetry needed for momentum conservation outside of the cold distribution limit we can fix the ``faulty'' part in expression (\ref{eq:DVDTloc}):
\[
\frac{4\pi}{3} \left(\frac{m_b}{2\pi k_BT_b}\right)^{3/2}R(u)
\]
by replacing throughout the quadratic thermal velocity of target particles with an expression which is symmetric in the exchange $a\leftrightarrow b$, for example the mean:
\[
\frac{k_BT_b}{m_b} \rightarrow \left(\frac{k_BT_b}{m_b}\right)^* = \frac{1}{n_a+n_b}\left(n_a\frac{k_BT_a}{m_a}+n_b\frac{k_BT_b}{m_b}\right)
\]
which is an easily accessible quantity in practice since it is the ratio of total pressure to total density, or the sum:
\begin{equation}
\frac{k_BT_b}{m_b} \rightarrow \left(\frac{k_BT_b}{m_b}\right)^* = \frac{k_BT_a}{m_a}+\frac{k_BT_b}{m_b} \label{eq:Tbstar}
\end{equation}
which is more satisfactory from a physical point of view because it is supposed to be the squared average relative velocity in the collision of a particle $a$ on a particle $b$.
Using the synthetic expression for $R(u)$ given above, the slowing-down rate is finally put in a form with the requested symmetry:
\begin{align*}
& \frac{dv_{a,i}}{dt} = -\frac{n_bm_b}{n_am_a+n_bm_b}\frac{v_{a,i}-v_{b,i}}{\tau_R} \\
& \frac{d}{dt}(v_{a,i}-v_{b,i}) = -\frac{v_{a,i}-v_{b,i}}{\tau_R}
\end{align*}
where the slowing-down time $\tau_R$ of ions by target particles reads, in the case of ions (labelled by subscript $b$):
\begin{align}
\tau_{Rab} &= \frac{m_a^2m_b^2}{4\pi e^4Z_a^2Z_b^2 (m_a+m_b)(n_am_a+n_bm_b)\mathrm{Log}\Lambda_{ab}} \nonumber \\
   & \hskip 0em \times\left(|\vec{v}_a-\vec{v}_b|^2 + \left(\frac{9\pi}{2}\right)^{1/3}\left(\frac{k_BT_b}{m_b}\right)^* \right)^{3/2} \label{eq:tauRab}
\end{align}
It may be questionable to use, in the limit of a vanishing relative velocity, an expression of the slowing-down rate which is strictly valid for a single particle $a$ colliding on target particles $b$.
In the limit $\vec{v}_a-\vec{v}_b\to 0$ the slowing-down rate can be computed exactly if the two distributions are assumed to remain Maxwellian (although this is questionable when $T_a\neq T_b$).

Thus if $f_a$ is the Maxwellian for particles of mass $m_a$ with parameters $n_a$, $\vec{v}_a$ and $T_a$, we can write, to first order in $|\vec{v}_a-\vec{v}_b|$:
\begin{align*}
f_a(\mathbf{c}) &\sim n_a\left(\frac{m_a}{2\pi k_BT_a}\right)^{3/2}e^{-m_a|\vec{c}-\vec{v}_b|^2/2k_BT_a} \\
 & \hskip 7em \times\left(1 + \frac{m_a}{k_BT_a}(v_{a,i}-v_{b,i})(c_i-v_{b,i})\right)
\end{align*}
Using expression (\ref{eq:dSdv}) we can then explicitly compute the slowing-down rate:
\[
\frac{d}{dt}(v_{a,i}-v_{b,i}) \sim -\frac{v_{a,i}-v_{b,i}}{\tau_{RM}}
\]
with the slowing-down time
\[
\tau_{RM} = \frac{3m_a^2m_b^2\left(\frac{k_BT_a}{m_a}+\frac{k_BT_b}{m_b}\right)^{3/2}}{4\sqrt{2\pi} e^4Z_a^2Z_b^2 (m_a+m_b)(n_am_a+n_bm_b) \mathrm{Log}\Lambda_{ab}} 
\]
The vanishing-velocity limit of the approximate expression (\ref{eq:tauRab}) is recovered provided the second definition of the average temperature (\ref{eq:Tbstar}) (the sum of the quadratic mean velocities) is used.

In the case of electrons, we only write the part corresponding to the time derivative on the ion velocity:
\[
\frac{dv_{ai}}{dt} = -\frac{v_{ai}}{\tau_{Rae}}
\]
since momentum conservation in collisions involves all ion species simultaneously, and the electron bulk velocity relaxes very quickly to a value determined by that set of species. We then get, to lowest order in powers of the electron/ion mass ratio:
\[
\tau_{Rae} \sim \frac{3m_a(k_BT_e)^{3/2}}{4\sqrt{2\pi}e^4Z_a^2m_e^{1/2}n_e\mathrm{Log}\Lambda_{ae}}
\]

\paragraph{Definition of a diffusion time.}
From the form of the second Fokker-Planck collision term, a velocity diffusion (or thermalization) time $\tau_D$ of ions by target particles can be defined in the following way:
\[
\frac{\partial f}{\partial t} = \frac{\partial}{\partial c_i}\left(D_{ij}\frac{\partial f}{\partial c_j}\right)
\]
with
\[
 \mbox{Tr}(\mathbf{D}) = \frac{1}{2}\frac{d\langle c^2\rangle}{dt} = \frac{\langle c^2\rangle}{\tau_D}
\]
where $\langle c^2\rangle$ is the mean quadratic width of the distribution $f$, which reads for a Maxwellian:
\[
\langle c^2\rangle = 3\frac{k_BT}{m}
\]
In the case of electrons this is:
\[
\tau_{Dae} = \frac{T_a}{T_e} \tau_{Rae}
\]
and for ions (labelled by subscript $b$):
\[
\tau_{Dab} = \frac{T_a}{T_b} \tau_{Rab} \frac{3R(u)}{R(u)+2L(u)}
\]
where
\[
u = \left(\frac{m_b}{2k_BT_b}\right)^{1/2}v 
\]
This also reads
\[
\tau_{Dab} = \frac{3m_ak_BT_a}{4\pi e^4Z_a^2Z_b^2n_b\mathrm{Log}\Lambda_{ab}}\frac{v}{\mathrm{erf}\left(\left(\frac{m_b}{2k_BT_b}\right)^{1/2}v\right)}
\]
A practical approximate formula for the ion-ion diffusion time, with the correct limits for $u\to 0$ and $u\to\infty$, can be designed in the same way as for the slowing-down time:
\begin{equation}
\tau_{Dab} = \frac{3m_ak_BT_a}{4\pi e^4Z_a^2Z_b^2n_b\mathrm{Log}\Lambda_{ab}}\left(v^2+\frac{\pi}{2}\frac{k_BT_b}{m_b}\right)^{1/2} \label{eq:tauDab}
\end{equation}
Close to thermal equilibrium, for all particle species a Fokker-Planck term is recovered, which takes on the form:
\[
\frac{\partial f_a}{\partial t} = \frac{1}{\tau_{Rab}}\frac{\partial}{\partial c_i}\left(c_if_a + \frac{k_BT_b}{m_a}\frac{\partial f_a}{\partial c_i}\right)
\]
with a characteristic time which is the slowing-down time $\tau_{Rab}$ of particles $a$ by the distribution of target particles $b$.

\paragraph{A global relaxation time.}
The collision times estimated in the preceding sections pertain to the evolution of localized parts of the test-particle distribution in velocity space.
Those times will acquire a global meaning for the whole distribution if they can be defined so as to keep, at least approximately, the same value over the region where the test distribution function is not negligible.
This clearly applies to the relaxation of an ion distribution on electrons, thanks to the large difference in characteristic velocities, or for ion-ion collisions in the case of a plasma interpenetration with a relative velocity larger than the ion thermal velocity.
Moreover, in those two cases, the self-collisions of the ion distribution draw it back to the Maxwellian, which strengthens the global character of the interaction with target particles.

In the case of ion-ion collisions with a relative velocity comparable with the thermal velocity, as already mentioned by Kogan at the end of his paper \cite{KOG611}, it is more difficult to define a global relaxation coefficient, even though it can be explicitly calculated in the case of two Maxwellians.
The result given by Kogan for temperature relaxation with a vanishing relative velocity (once corrected for a missing factor with respect to the Rosenbluth collision term \cite{ROS573}), is:
\begin{equation}
\tau_{ab} = \frac{3(m_bk_BT_a+m_ak_BT_b)^{3/2}}{8\sqrt{2\pi m_am_b}e^4Z_a^2Z_b^2n_b\mathrm{Log}\Lambda_{ab}} \label{eq:tauab}
\end{equation}
which is the characteristic time to use in the temperature relaxation equation:
\[
\frac{dT_a}{dt} = \frac{T_b-T_a}{\tau_{ab}}
\]
which leads to the symmetric rate
\[
\frac{d}{dt}(T_a-T_b) = \frac{T_b-T_a}{\tau_{TM}}
\]
where
\begin{align*}
\tau_{TM} &= \left(\frac{1}{\tau_{ab}}+\frac{1}{\tau_{ba}}\right)^{-1} \\
   &= \frac{3m_am_b\left(\frac{k_BT_a}{m_a}+\frac{k_BT_b}{m_b}\right)^{3/2}}{8\sqrt{2\pi} e^4Z_a^2Z_b^2 (n_a+n_b) \mathrm{Log}\Lambda_{ab}}
\end{align*}
This time is very similar to the limit slowing-down time $\tau_{RM}$, and is actually the same in the case of equal-mass particles $m_a=m_b$.
This seems to make expression (\ref{eq:tauRab}) a decent candidate for a global relaxation time taking into account plasma interpenetration and/or pressure anisotropy.

But actually the faster particles in the distribution will relax more slowly, so that the distribution will be distorted away from the Maxwellian.
In particular when the temperatures are very different ($T_a\gg T_b$) we know \cite{LAR071} that the test-particle distribution will acquire a colder component in the target particle region, while the rest of the distribution will slow down with an almost vanishing-divergence current in velocity space, which can actually be used to model the slowing-down of the fast $\alpha$ particles from fusion reactions in ICF \cite{PEI14A,PEI143}.
Even when the temperatures are the same, as discussed in the text, in actual interpenetration calculations performed with a kinetic code, it is found (see the animation provided as supplementary material) that at the end of the relaxation the distribution shifts from the two-beam to a bi-Maxwellian shape, and accordingly the limit relaxation rate near isotropy is closer to the analytic value found in the latter case (see Appendix \ref{app:bimax}).
Thus obviously, instead of looking for a single analytic formula valid for all cases, we have to design a heuristic relaxation rate accounting for the actual behaviour of the plasma, supposedly found in kinetic calculations.

Of course we should not expect a relaxation rate, however cleverly designed, to account for the diversity of kinetic effects.
It will only be used as a reasonable order of magnitude in the situations expected to occur in hohlraum plasmas, and specifically as an important input in their modeling through extended hydrodynamics.

\section{Relaxation of the anisotropy of a bi-Maxwellian}
\label{app:bimax}

Kogan \cite{KOG611} has given an analytic expression of the rate of self-collision relaxation of a bi-Maxwellian, i.e. a distribution reading:
\begin{align}
f(c_x,c_\bot) &= N\left(\frac{m}{2\pi k_BT_\parallel}\right)^{1/2}\frac{m}{2\pi k_BT_\bot} \nonumber \\
   & \hskip 4em \times\mathrm{exp}\left(\frac{-m}{2k_B}\left(\frac{c_x^2}{T_\parallel}+\frac{c_\bot^2}{T_\bot}\right)\right) \label{eq:fbimax}
\end{align}
The general result, valid for all values of the degree of anisotropy, is the following (a correction factor 2 was included, bringing Kogan's expression of the collision term in agreement with that of Rosenbluth \textit{et al.} \cite{ROS573}):
\begin{align}
\frac{dT_\parallel}{dt} &= \frac{8Z^4e^4N\mbox{Log}\Lambda}{5}\left(\frac{\pi}{m (k_BT)^3}\right)^{1/2} \nonumber \\
   & \hskip 9em \times (T-T_\parallel)F(\frac{T_\parallel-T}{T}) \label{eq:Kogan}
\end{align}
where the function $F(x)$ reads:
\begin{align}
F(x) &= -\frac{5\sqrt{1+x}}{x^2} \left[\rule{0pt}{2em}1 + \frac{1}{\sqrt{6}}\left(\frac{\sqrt{x}}{2\sqrt{1+x}}\right.\right. \nonumber \\
   & \hskip 3em \left.\left.- \frac{\sqrt{1+x}}{\sqrt{x}}\right) \mbox{Log}\frac{\sqrt{1+x}+\sqrt{\frac{3}{2}x}}{\sqrt{1+x}-\sqrt{\frac{3}{2}x}}\right]  \label{eq:FKogan}
\end{align}
That expression is valid without restrictions for $x>0$ ($T_\parallel>T_\bot$), and in the reverse case its analytic extension in the complex plane of values of $\sqrt{x}$ must be used, noticing that for all complex values of $z$
\[
\mbox{Log}\frac{1+iz}{1-iz} = 2i\mbox{Arctg}z
\]
so that for $x<0$~:
\begin{align*}
F(x) &= -\frac{5\sqrt{1+x}}{x^2} \left[\rule{0pt}{2em}1 - \frac{1}{\sqrt{6}}\left(\frac{\sqrt{-x}}{\sqrt{1+x}}\right.\right. \\
   &  \hskip 7em \left.\left. +2\frac{\sqrt{1+x}}{\sqrt{-x}}\right) \mbox{Arctg}\frac{\sqrt{-\frac{3}{2}x}}{\sqrt{1+x}}\right]
\end{align*}
$F(x)$ is plotted as the dashed green curve on Fig. 5.
For a small anisotropy $F(x)\buildlow{\rightarrow}{0}1$, leading to the following definition of the ion-ion collision time for a near-Maxwellian distribution:
\[
\frac{dT_\parallel}{dt} = \frac{T-T_\parallel}{\tau_{Max}}
\]
with
\begin{equation}
\tau_{Max} = \frac{5m_i^{1/2}(k_BT_i)^{3/2}}{8\sqrt{\pi}Z^4e^4N_i\mbox{Log}\Lambda_{ii}} \label{eq:tauMbimax}
\end{equation}
The particular numerical factor in the above expression of the collision time arises from the expansion of the relaxation rate about isotropy ($x\sim 0$), as can be cross-checked through a direct calculation from the Rosenbluth potentials, given in the next paragraph.
It is specific to the relaxation of the anisotropy of a bi-Maxwel\-lian distribution, and numerically different from those found in other collisional relaxation processes near iso\-tropy, even though its order of magnitude and functional dependencies on mass, density and temperature are the same.

\paragraph{Direct calculation from the Rosenbluth potentials.}

The evolution of the second-order moment of the distribution due to collisions reads:
\begin{align*}
\frac{d}{dt}(Nk_BT_\parallel) &= \int mc_x^2\left(\frac{\partial f}{\partial t}\right)_c(\vec{c})d^3c \\
 &= 8\pi m\Gamma\int\left(\mathcal{S}\frac{\partial}{\partial c_x}(2c_xf) + \frac{\partial\mathcal{T}}{\partial c_x}\frac{\partial f}{\partial c_x}\right)d^3c
\end{align*}
where expressions (\ref{eq:dfdtcab})-(\ref{eq:Rpots}) were inserted, dropping all species-specific subscripts since a single species is involved. Using the integral expressions of the potentials (\ref{eq:Sb}) and (\ref{eq:Tb}), this reads:
\begin{align*}
\frac{d}{dt}(Nk_BT_\parallel) &= -m\Gamma\int\kern -1.5ex\int\frac{f(\vec{c}^\prime)}{|\vec{c}-\vec{c}^\prime|} \left(\rule{0pt}{1.35em}4f(\vec{c})\right. \\
  & \hskip 5em \left. +(5c_x-c_{x}^\prime)\frac{\partial f}{\partial c_x}(\vec{c})\right)d^3cd^3c^\prime
\end{align*}
Using Kogan's change of variables (with unit Jacobian):
\[
(\vec{c},\vec{c}^\prime) \rightarrow \left(\vec{u} = \vec{c}-\vec{c}^\prime\enspace,\enspace\vec{t} = \frac{\vec{c}+\vec{c}^\prime}{2}\right)
\]
and taking into account that $f$ is the bi-Maxwellian (\ref{eq:fbimax}) to write its $c_x$-derivative, we obtain:
\begin{align*}
\frac{d}{dt}(Nk_BT_\parallel) &= -4m\Gamma\int\kern -1.5ex\int\left(1 - \frac{m}{k_BT_\parallel}\left(t_x+\frac{3}{4}u_x\right)\right. \\
   & \hskip -1em \left. \times\left(t_x+\frac{u_x}{2}\right)\rule{0pt}{1.35em}\right) f\left(\mathbf{t}-\frac{\mathbf{u}}{2}\right) f\left(\mathbf{t}+\frac{\mathbf{u}}{2}\right)\frac{d^3u}{|\vec{u}|}d^3t
\end{align*}
or, inserting the expression of the distribution function:
\begin{align*}
\frac{d}{dt}(k_BT_\parallel) &= -4m\Gamma N\left(\frac{m}{2\pi k_BT_\bot}\right)^2\frac{m}{2\pi k_BT_\parallel} \\
   & \hskip 1em\times\int\kern -1.5ex\int\left(1 - \frac{m}{k_BT_\parallel}\left(t_x+\frac{3}{4}u_x\right)\left(t_x+\frac{u_x}{2}\right)\right) \\
 & \hskip 3em \times\mathrm{exp}\left(\frac{-m}{k_B}\left(\frac{t_x^2}{T_\parallel}+\frac{t_\bot^2}{T_\bot}\right)\right) \\
 & \hskip 4em \times\mathrm{exp}\left(\frac{-m}{4k_B}\left(\frac{u_x^2}{T_\parallel}+\frac{u_\bot^2}{T_\bot}\right)\right)\frac{d^3u}{|\vec{u}|}d^3t
\end{align*}
Integrations over $\vec{t}$ are straightforward, and we are left with the following integral over $\vec{u}$:
\begin{align*}
\frac{d}{dt}(k_BT_\parallel) &= -2m\Gamma N\frac{m}{4\pi k_BT_\bot} \left(\frac{m}{4\pi k_BT_\parallel}\right)^{1/2} \\
   & \hskip -2em \times \int\left(1 - \frac{3mu_x^2}{4k_BT_\parallel}\right) \mathrm{exp}\left(\frac{-m}{4k_B}\left(\frac{u_x^2}{T_\parallel}+\frac{u_\bot^2}{T_\bot}\right)\right)\frac{d^3u}{|\vec{u}|}
\end{align*}
We now define:
\[
\left(\frac{m}{4k_BT_\parallel}\right)^{1/2}u_x = r\cos\theta
\]
and
\[
\left(\frac{m}{4k_BT_\bot}\right)^{1/2}u_\bot = r\sin\theta
\]
which splits the integral into an angular part and a radial part which can be integrated in a straightforward way, finally leading to:
\begin{align*}
\frac{d}{dt}(k_BT_\parallel) &= -\frac{m\Gamma N}{\sqrt{\pi}}\int_0^\pi\frac{(1-3\cos^2\theta)\sin\theta d\theta}{\left(\frac{k_BT_\parallel}{m}\cos^2\theta+\frac{k_BT_\bot}{m}\sin^2\theta\right)^{1/2}} \\
   &= \frac{8\sqrt{\pi}Z^4e^4N\mbox{Log}\Lambda}{\sqrt{mk_BT}}\frac{\sqrt{1+x}}{x} \times \\
   & \hskip -4em \left[1 - \frac{1}{\sqrt{6}}\left(\sqrt{\frac{1+x}{x}}-\frac{1}{2}\sqrt{\frac{x}{1+x}}\right)\mathrm{Log}\frac{\sqrt{1+x}+\sqrt{\frac{3x}{2}}}{\sqrt{1+x}-\sqrt{\frac{3x}{2}}}\right]
\end{align*}
where we inserted $x=(T_\parallel-T)/T$.
It can be checked that the expression (\ref{eq:Kogan})-(\ref{eq:FKogan}) obtained by Kogan is recovered (including the previously mentioned correction factor).
Expanding the above expression about isotropy ($x\sim 0$) we find
\begin{align*}
\frac{d}{dt}(k_BT_\parallel) & \buildlow{\sim}{x\sim 0} -\frac{8\sqrt{\pi}Z^4e^4N\mbox{Log}\Lambda}{5\sqrt{mk_BT}}x \\
     & \buildlow{\sim}{x\sim 0} \frac{8\sqrt{\pi}Z^4e^4N\mbox{Log}\Lambda}{5\sqrt{m(k_BT)^3}}(k_BT-k_BT_\parallel)
\end{align*}

\section{Moments of a two-component distribution with azimuthal symmetry}
\label{app:moms}

The velocity distribution function is assumed to be in the form $f(\vec{c}) = f^{(1)}(\vec{c}) + f^{(2)}(\vec{c})$ with
\[
f^{(n)}(\vec{c}) = \frac{\rho_n}{m}f_\parallel^{(n)}(\vec{c})f_\bot^{(n)}(\vec{c})
\]
and the factors are defined by
\begin{multline}
f_\parallel^{(n)}(\vec{c}) = \left(\frac{m}{2k_BT_{\parallel n}}\right)^{1/2} \times \\
   F_\parallel\left(\left(\frac{m}{2k_BT_{\parallel n}}\right)^{1/2}\Omega_i(c_i-v_{ni})\right)
\label{eq:Cfparn}
\end{multline}
and
\begin{multline}
f_\bot^{(n)}(\vec{c}) = \frac{m}{2k_BT_{\bot n}}F_\bot\left((c_i-v_{ni})\times\rule{0pt}{1.3em}\right. \\
    \left.\left[\frac{m}{2k_BT_{\bot n}}(\delta_{ij}-\Omega_i\Omega_j)\right](c_j-v_{nj})\right)
\label{eq:Cfperpn}
\end{multline}
$T_{\parallel n}$ and $T_{\bot n}$ are the parallel and perpendicular temperatures of beam number $n$, $\vec{\Omega}=(\vec{v}_2-\vec{v}_1)/|\vec{v}_2-\vec{v}_1|$, and the functions $F_\parallel$ and $F_\bot$ are normalized as follows:
\[
\int_{-\infty}^\infty F_\parallel(x)dx = 1 \quad ; \quad \int_0^\infty F_\bot(x)dx = \frac{1}{\pi}
\]
and for higher moments:
\begin{align*}
& \int_{-\infty}^\infty xF_\parallel(x)dx = 0 \quad ; \quad \int_{-\infty}^\infty x^2F_\parallel(x)dx = \frac{1}{2} \\
& \int_0^\infty xF_\bot(x)dx = \frac{1}{\pi}
\end{align*}
In the specific case of a bi-Maxwel\-lian distribution, we have:
\[
F_\parallel(x) = \frac{1}{\sqrt{\pi}}\mathrm{e}^{-x^2} \quad \mathrm{and} \quad F_\bot(x) = \frac{1}{\pi}\mathrm{e}^{-x}
\]
The first two moments are obviously
\begin{align*}
\rho &= m\int \left(f^{(1)}(\vec{c}) + f^{(2)}(\vec{c})\right)d^3c = \rho_1+\rho_2 \\
\rho \vec{v} &= m\int \vec{c}\left(f^{(1)}(\vec{c}) + f^{(2)}(\vec{c})\right)d^3c = \rho_1\vec{v}_1+\rho_2\vec{v}_2
\end{align*}
The bulk velocity is the barycentre of the distribution:
\[
\vec{v} = \frac{\rho_1\vec{v}_1+\rho_2\vec{v}_2}{\rho_1+\rho_2}
\]
The moment of order 2 reads, using notations from \cite{VIK114}:
\[
mM_{ij}^2 = m\int c_ic_jf(\vec{c})d^3c = \rho v_iv_j+P_{ij}
\]
where $P_{ij}$ is the pressure tensor:
\begin{align}
P_{ij} &= m\int (c_i-v_i)(c_j-v_j)\left(f^{(1)}(\vec{c})+f^{(2)}(\vec{c})\right)d^3c \nonumber \\
       &= P_{ij}^{(1)}+P_{ij}^{(2)} + \frac{\rho_1\rho_2}{\rho_1+\rho_2}|\vec{v}_2-\vec{v}_1|^2\Omega_i\Omega_j \label{eq:apij}
\end{align}
where
\[
P_{ij}^{(n)} = m\int (c_i-v_{ni})(c_j-v_{nj})f^{(n)}(\vec{c})d^3c
\]
The vector $\vec{\Omega}$ being given, an orthonormal basis\linebreak$(\vec{\Omega},\vec{U},\vec{V})$ can be defined, in which the vector $\vec{x}=\vec{c}-\vec{v}_n$ has components $(x_\parallel,x_U,x_V)$. The pressure tensor of component $n$ then reads
\begin{multline*}
P_{ij}^{(n)} = \int(x_\parallel^2\Omega_i\Omega_j+x_U^2U_iU_j+x_V^2V_iV_j + ...)\times \\
             \rule{2em}{0pt} \frac{\rho_nm^{3/2}}{(2k_BT_{\parallel n})^{1/2}2k_BT_{\bot n}}
             F_\parallel\left(\left(\frac{m}{2k_BT_{\parallel n}}\right)^{1/2}x_\parallel\right)\times \\
              F_\bot\left(-\left[\frac{m(x_U^2+x_V^2)}{2k_BT_{\bot n}}\right]\right) d^3x
\end{multline*}
where the ellipsis stands for crossed terms such as \linebreak
$x_\parallel x_U\Omega_iU_j$ whose integral against $F_\bot$ vanishes. The sum of non-vanishing terms is
\begin{align*}
P_{ij}^{(n)} &= \rho_n\frac{k_BT_{\parallel n}}{m}\Omega_i\Omega_j + \rho_n\frac{k_BT_{\bot n}}{m}(U_iU_j+V_iV_j) \\
             &= \rho_n\frac{k_BT_{\parallel n}}{m}\Omega_i\Omega_j + \rho_n\frac{k_BT_{\bot n}}{m}(\delta_{ij}-\Omega_i\Omega_j)
\end{align*}
Substituting that expression into (\ref{eq:apij}) we finally get the pressure tensor for a two-component distribution with azimuthal symmetry around the axis $\vec{\Omega}$:
\[
P_{ij} = \Ppa \Omega_i\Omega_j + \Ppe (\delta_{ij}-\Omega_i\Omega_j)
\]
with
\begin{align*}
\Ppa &= \rho_1\frac{k_BT_{\parallel 1}}{m}+\rho_2\frac{k_BT_{\parallel 2}}{m} +\frac{\rho_1\rho_2}{\rho_1+\rho_2}|\vec{v}_2-\vec{v}_1|^2 \\
\Ppe &= \rho_1\frac{k_BT_{\bot 1}}{m}+\rho_2\frac{k_BT_{\bot 2}}{m}
\end{align*}
The moment of order 3 reads, using notations from \cite{VIK114}:
\begin{align*}
mM_{ijk}^3 &= m\int c_ic_jc_kf(\vec{c})d^3c \\
           &= \rho v_iv_jv_k+v_iP_{jk}+v_jP_{ik}+v_kP_{ij}+Q_{ijk}
\end{align*}
where $Q_{ijk}$ is twice the heat flux tensor:
\begin{align*}
Q_{ijk} &= m\int (c_i-v_i)(c_j-v_j)(c_k-v_k) \times \\
        & \hskip 7em \left(f^{(1)}(\vec{c})+f^{(2)}(\vec{c})\right)d^3c \\
        &= Q_{ijk}^{(1)}+Q_{ijk}^{(2)} \\
        &  +(v_{1i}-v_i)P_{jk}^{(1)}+(v_{1j}-v_j)P_{ik}^{(1)}+(v_{1k}-v_k)P_{ij}^{(1)} \\
        &  +(v_{2i}-v_i)P_{jk}^{(2)}+(v_{2j}-v_j)P_{ik}^{(2)}+(v_{2k}-v_k)P_{ij}^{(2)} \\
        &  +\frac{\rho_1\rho_2(\rho_1-\rho_2)}{(\rho_1+\rho_2)^2}|\vec{v}_2-\vec{v}_1|^3\Omega_i\Omega_j\Omega_k
\end{align*}
$Q_{ijk}^{(n)}$ is the moment of order 3 restricted to component $n$ and computed in the reference frame centered on its bulk velocity $\vec{v}_{n}$:
\[
Q_{ijk}^{(n)} = m\int (c_i-v_{ni})(c_j-v_{nj})(c_k-v_{nk})f^{(n)}(\vec{c})d^3c
\]
Assuming $\int_{-\infty}^\infty x^3F_\parallel(x)dx=0$, $Q_{ijk}^{(n)}$ can be computed in the same way as $P_{ij}^{(n)}$ above.
As expected, the result vanishes since the integral contains only terms in which at least one of the factors $x_p$ enters to an odd power.
Inserting values already obtained for $v_{ni}-v_i$ and $P_{ij}^{(n)}$ and considering the dependence of the result on degrees of freedom parallel and perpendicular to $\vec{\Omega}$, we finally get
\begin{multline*}
Q_{ijk} = \Qpa \Omega_i\Omega_j\Omega_k + \Qpe [\Omega_i(\delta_{jk}-\Omega_j\Omega_k) \\
         +\Omega_j(\delta_{ik}-\Omega_i\Omega_k)+\Omega_k(\delta_{ij}-\Omega_i\Omega_j)]
\end{multline*}
with
\begin{align*}
\Qpa &= \frac{\rho_1\rho_2}{\rho_1+\rho_2}|\vec{v}_2-\vec{v}_1|\left[3\left(\frac{k_BT_{\parallel 2}}{m}-\frac{k_BT_{\parallel 1}}{m}\right)\right. \\
        & \hskip 11em \left. +\frac{\rho_1-\rho_2}{\rho_1+\rho_2}|\vec{v}_2-\vec{v}_1|^2\right] \\
\Qpe &= \frac{\rho_1\rho_2}{\rho_1+\rho_2}|\vec{v}_2-\vec{v}_1|\left(\frac{k_BT_{\bot 2}}{m}-\frac{k_BT_{\bot 1}}{m}\right)
\end{align*}
The heat flux vector is
\[
q_i=\frac{1}{2}Q_{ijj} = \frac{1}{2}(\Qpa+2\Qpe)\Omega_i
\]
The intrinsic moment of order 4 (computed in the reference frame centered on the global bulk velocity $\vec{v}$) reads
\begin{align*}
R_{ijkm} &= m\int(c_i-v_i)(...)\left(f^{(1)}(\vec{c})+f^{(2)}(\vec{c})\right)d^3c \\
         &= R_{ijkm}^{(1)} +\left[(v_{1i}-v_i)(v_{1j}-v_j)P_{km}^{(1)} + ...\ \right] \\
         & \ +\rho_1(v_{1i}-v_i)(v_{1j}-v_j)(v_{1k}-v_k)(v_{1m}-v_m) \\
         & \ +R_{ijkm}^{(2)} +\left[(v_{2i}-v_i)(v_{2j}-v_j)P_{km}^{(2)} + ...\ \right] \\
         & \ +\rho_2(v_{2i}-v_i)(v_{2j}-v_j)(v_{2k}-v_k)(v_{2m}-v_m)
\end{align*}
where the ellipsis stands for permutations making the preceding expression symmetric, and $R_{ijkm}^{(n)}$ is the moment of order 4 restricted to component $n$ and computed in the reference frame centered on its bulk velocity $\vec{v}_{n}$:
\[
R_{ijkm}^{(n)} = m\int x_ix_jx_kx_mf^{(n)}(\vec{x})d^3x
\]
where the same definition as above $x_i=c_i-v_{ni}=x_\parallel\Omega_i+x_UU_i+x_VV_i$ is used. Hence,
\begin{align*}
R_{ijkm}^{(n)} &= m\int x_\parallel^4f^{(n)}(\vec{x})d^3x\ \Omega_i\Omega_j\Omega_k\Omega_m \\
               & \hskip -2em + m\int x_\parallel^2x_U^2f^{(n)}(\vec{x})d^3x \left[\Omega_i\Omega_j\left(\delta_{km}-\Omega_k\Omega_m\right) + ...\ \right] \\
               &+ m\int x_U^2x_V^2f^{(n)}(\vec{x})d^3x \left[U_iU_jV_kV_m +...\ \right] \\
               &+ m\int x_U^4f^{(n)}(\vec{x})d^3x \left(U_iU_jU_kU_m + V_iV_jV_kV_m\right)
\end{align*}
In the latter, it was noticed that
\[
\int x_U^{2p}f^{(n)}(\vec{x})d^3x = \int x_V^{2p}f^{(n)}(\vec{x})d^3x
\]
since the distribution is assumed isotropic in the transverse velocity plane. After some manipulations we get
\begin{align*}
\left[U_iU_jV_kV_m +...\ \right] &= \left(\delta_{ij}-\Omega_i\Omega_j\right)\left(\delta_{km}-\Omega_k\Omega_m\right) \\
   &\quad + \left(\delta_{ik}-\Omega_i\Omega_k\right)\left(\delta_{jm}-\Omega_j\Omega_m\right) \\
   &\quad + \left(\delta_{im}-\Omega_i\Omega_m\right)\left(\delta_{jk}-\Omega_j\Omega_k\right) \\
   &\quad -3\left(U_iU_jU_kU_m + V_iV_jV_kV_m\right)
\end{align*}
and since, once again due to isotropy in the transverse velocity plane,
\[
\int x_U^4f^{(n)}(\vec{x})d^3x = 3\int x_U^2x_V^2f^{(n)}(\vec{x})d^3x
\]
the expression for $R_{ijkm}^{(n)}$ can be simplified somewhat:
\begin{align*}
& R_{ijkm}^{(n)} = m\int x_\parallel^4f^{(n)}(\vec{x})d^3x\ \Omega_i\Omega_j\Omega_k\Omega_m \\
& \hskip 1.5em + m\int x_\parallel^2x_U^2f^{(n)}(\vec{x})d^3x \left[\Omega_i\Omega_j\left(\delta_{km}-\Omega_k\Omega_m\right) + ...\ \right] \\
& \hskip 2em + m\int x_U^4f^{(n)}(\vec{x})d^3x \ \times \\
& \hskip 7em \frac{1}{3}\left[\left(\delta_{ij}-\Omega_i\Omega_j\right)\left(\delta_{km}-\Omega_k\Omega_m\right)\right. \\
& \hskip 8em + \left(\delta_{ik}-\Omega_i\Omega_k\right)\left(\delta_{jm}-\Omega_j\Omega_m\right) \\
& \hskip 9em\left. + \left(\delta_{im}-\Omega_i\Omega_m\right)\left(\delta_{jk}-\Omega_j\Omega_k\right)\right]
\end{align*}
Inserting the form chosen for the distribution components, we get
\begin{align*}
m\int x_\parallel^4f^{(n)}(\vec{x})d^3x &= R_{\parallel\parallel}^{(n)} \\
m\int x_\parallel^2x_U^2f^{(n)}(\vec{x})d^3x &= \rho_n\frac{k_BT_{\parallel n}}{m}\frac{k_BT_{\bot n}}{m} \\
m\int x_U^4f^{(n)}(\vec{x})d^3x &= 3\rho_n\left(\frac{k_BT_{\bot n}}{m}\right)^2 
\end{align*}
where, in the specific cases of a bi-Maxwellian or a ``waterbag'' (flat-top) distribution, respectively:
\[
R_{\parallel\parallel}^{(n)} = 3\rho_n\left(\frac{k_BT_{\parallel n}}{m}\right)^2 \quad\mathrm{or}\quad \frac{9}{5}\rho_n\left(\frac{k_BT_{\parallel n}}{m}\right)^2
\]
The final expression for the tensor of order 4 is thus
\begin{multline*}
R_{ijkl} = R_{\parallel\parallel} \Omega_i\Omega_j\Omega_k\Omega_l
  + R_{\parallel\bot}[\Omega_i\Omega_j(\delta_{kl}-\Omega_k\Omega_l) + ...\ ] \\
      +R_{\bot\bot} [(\delta_{ij}-\Omega_i\Omega_j)(\delta_{kl}-\Omega_k\Omega_l) +...\ ]
\end{multline*}
where the first symmetrized bracket contains 6 terms, and the second one 3 terms. The components are
\begin{multline}
R_{\parallel\parallel} = R_{\parallel\parallel}^{(1)} + \rho_1\left[6\frac{k_BT_{\parallel 1}}{m}\frac{\rho_2^2|\vec{v}_2-\vec{v}_1|^2}{(\rho_1+\rho_2)^2}
      + \frac{\rho_2^4|\vec{v}_2-\vec{v}_1|^4}{(\rho_1+\rho_2)^4}\right] + \\
          R_{\parallel\parallel}^{(2)} + \rho_2\left[6\frac{k_BT_{\parallel 2}}{m}\frac{\rho_1^2|\vec{v}_2-\vec{v}_1|^2}{(\rho_1+\rho_2)^2}
      + \frac{\rho_1^4|\vec{v}_2-\vec{v}_1|^4}{(\rho_1+\rho_2)^4}\right] \label{eq:Rpapa2}
\end{multline}
\begin{multline}
R_{\parallel\bot} = \rho_1\frac{k_BT_{\bot 1}}{m}\left(\frac{k_BT_{\parallel 1}}{m}+\frac{\rho_2^2|\vec{v}_2-\vec{v}_1|^2}{(\rho_1+\rho_2)^2}\right) \\
         + \rho_2\frac{k_BT_{\bot 2}}{m}\left(\frac{k_BT_{\parallel 2}}{m}+\frac{\rho_1^2|\vec{v}_2-\vec{v}_1|^2}{(\rho_1+\rho_2)^2}\right) \label{eq:Rpape2}
\end{multline}
\begin{equation}
R_{\bot\bot} = \rho_1\left(\frac{k_BT_{\bot 1}}{m}\right)^2+\rho_2\left(\frac{k_BT_{\bot 2}}{m}\right)^2 \label{eq:Rpepe}
\end{equation}

\end{multicols}
\end{document}